\documentclass[pre,aps,floats,twocolumn,superscriptaddress,floatfix]{revtex4}
\usepackage{graphicx,amssymb,amsmath}
\begin{document}
\title{Thermodynamics of statistically interacting quantum gas in
  $\mathcal{D}$ dimensions}  
\author{Geoffrey G. Potter}
\affiliation{
  Department of Physics,
  University of Rhode Island,
  Kingston RI 02881, USA
}
\author{Gerhard M{\"{u}}ller}
\affiliation{
  Department of Physics,
  University of Rhode Island,
  Kingston RI 02881, USA
}
\email[Gerhard Müller]{gmuller@uri.edu}
\author{Michael Karbach}
\affiliation{
  Bergische Universit{\"{a}}t Wuppertal,
  Fachbereich Mathematik und Naturwissenschaften,
  D-42097 Wuppertal, Germany}
\affiliation{
  Department of Physics,
  University of Rhode Island,
  Kingston RI 02881, USA
}
\email[Michael Karbach]{michael@karbach.org}
\pacs{75.10.-b}
\begin{abstract}
  We present the exact thermodynamics (isochores, isotherms, isobars, response
  functions) of a statistically interacting quantum gas in $\mathcal{D}$
  dimensions. The results in $\mathcal{D}=1$ are those of the thermodynamic
  Bethe ansatz for the nonlinear Schr{\"{o}}dinger model, a gas with repulsive
  two-body contact potential. In all dimensions the ideal boson and fermion
  gases are recovered in the weak-coupling and strong-coupling limits,
  respectively. For all nonzero couplings ideal fermion gas behavior emerges
  for $\mathcal{D}\gg1$ and, in the limit $\mathcal{D}\to\infty$, a phase
  transition occurs at $T>0$. Significant deviations from ideal quantum gas
  behavior are found for intermediate coupling and finite $\mathcal{D}$.
\end{abstract}

\maketitle

%
\section{Introduction}\label{sec:intro}
%
The wave of experimental studies that led to the first observations of
Bose-Einstein condensation (BEC) and the development of measurement and
confinement technologies have renewed strong interest in the statistical
mechanics of interacting quantum gases \cite{PS02, PS03, GBM+01, GVL+01,
  SKC+01, PGS04}. This line of research can make good use of explicit
high-accuracy results from any type of analysis that goes beyond
low-density/high-temperature expansions and beyond mean-field theory. Of
particular interest are results for response functions, the very quantities
most directly amenable to experimental investigations.

Such results can be produced on a rigorous basis for quantum gases with
statistical interaction under very general circumstances as shown by Wu
\cite{Wu94, Wu95a}.  The concept of \emph{statistical} interaction introduced
by Haldane \cite{Hald91a} has proven to be a very useful methodological device
to capture the statistical mechanical properties of degrees of freedom subject
to \emph{dynamical} interaction. For several model systems in dimension
$\mathcal{D}=1$ the coupling between degrees of freedom can be substituted by a
generalized Pauli principle with no loss of rigor regarding the thermodynamic
analysis \cite{Hald91, Hald91a, Hald94, BW94, AKMW06}.

Whereas an equivalence between dynamical and statistical interaction is not
likely to be realized in $\mathcal{D}>1$ (apart from highly contrived
scenarios), models of statistically interacting degrees of freedom can stand on
their own. Their thermodynamic properties can be analyzed exactly in any
dimension $\mathcal{D}$, producing a full and consistent account of
fluctuations as will be demonstrated in this work \cite{fn1}. The exact results
emerging from this analysis make it possible to connect features of the
statistical interaction with features of a corresponding dynamical interaction.
The systematic study of such connections, in turn, opens the door to the design
of (exactly solvable) models of statistical interaction for the description of
thermodynamic phenomena associated with specific aspects of dynamical
interaction.

In a previous paper \cite{PMK07} we have established a benchmark in that regard
by exploring the thermodynamics of an ideal quantum gas with fractional
statistics in $\mathcal{D}$ dimensions -- a thermodynamic generalization of the
Calogero-Sutherland (CS) model \cite{Calo71, Suth71, Suth72} -- taking
advantage of techniques and results reported in previous studies \cite{Wu94,
  BW94, Isak94, IAMP96, JSSB96, MS99, Suth71, NW94, Wu95a, Iguc97a, Iguc97b,
  Iguc97c, IA00, Aoya01, Medv97, May64, VRH95, Lee97, SC04, Angh02}. In that
case the statistical interaction was limited to pairs of particles with
identical momenta.

Here we relax that constraint and consider a model system, again in
$\mathcal{D}$ dimensions, with a statistical interaction that extends to pairs
of particles with arbitrary momenta, a system moreover, whose statistical
interaction in $\mathcal{D}=1$ is equivalent to the dynamical interaction of a
model that is solvable (beyond thermodynamics) via Bethe ansatz: the nonlinear
Schr{\"{o}}dinger (NLS) model \cite{LL63, Lieb63, YY69, Yang70, KBI93, Taka99,
  Suth04}.

In Sec.~\ref{sec:statint} we review the concept of statistical interaction and
its use in statistical mechanics. We introduce the NLS model and the
generalization of its thermodynamics to $\mathcal{D}>1$. In
Sec.~\ref{sec:thenls} we describe the method of thermodynamic analysis applied
to the generalized NLS model. In Sec.~\ref{sec:selres} we discuss selected
thermodynamic properties thus calculated. In Sec.~\ref{sec:concl} we assess the
results in relation to existing benchmarks for ideal quantum gases with
fractional statistics.


%
\section{Statistical interaction}\label{sec:statint}  
%
The statistical interaction of any given model system is specified by a
generalized Pauli principle \cite{Hald91a}, expressing how the number of
states available to one particle is affected by the presence of other
particles:
\begin{equation}\label{eq:core}
  \Delta d_i \doteq -\sum_j g_{ij}\Delta n_j.
\end{equation}
The indices $i,j$ refer to particle species and the $g_{ij}$ are {\em
  statistical interaction coefficients}. For bosons we have $g_{ij}=0$ and for
fermions $g_{ij}=\delta_{ij}$.  Integrating Eq.~(\ref{eq:core}) yields the
holding capacity for particles of species $i$ in the presence of a specific
number of particles from each species:
\begin{equation}\label{eq:d1p}
  d_i = A_i -\sum_j g_{ij}(n_j-\delta_{ij}),
\end{equation}
where $A_i$ are {\em statistical capacity constants}. The number of many-particle
states composed of $\{n_i\}$ statistically interacting particles is 
\begin{equation}\label{eq:prodWa}
  W(\{n_i\})=\prod_i \left( \begin{tabular}{c} $d_i+n_i-1$ \\ $n_i$ 
\end{tabular}\right).
\end{equation}

The three principal specifications of a system of particles subject to a
statistical interaction are sets of (i) energies $\epsilon_i$, (ii) capacity
constants $A_i$, and (iii) interaction coefficients $g_{ij}$. The grand
potential of such a system can be expressed in the form \cite{Wu94}
\begin{equation}
  \label{eq:71}
  \Omega=-k_BT\sum_iA_i\ln\left[\frac{1+w_i}{w_i}\right],
\end{equation}
where the $w_i$ are determined by the nonlinear algebraic
equations,
\begin{equation}
  \label{eq:54}
  \frac{\epsilon_i-\mu}{k_BT} = 
\ln(1+w_i)-\sum_jg_{ji}\ln\left(\frac{1+w_j}{w_j}\right).
\end{equation}
The control variables are $T$ (temperature) and $\mu$ (chemical potential).
The average numbers of particles, $\langle n_i\rangle$, are related to the
$w_i$ by the linear equations,
\begin{equation}
  \label{eq:1}
  \sum_j\left(\delta_{ij}w_j+g_{ij}\right)\langle n_j\rangle = A_i.
\end{equation}
If $g_{ij}=g_i\delta_{ij}$ then all Eqs.~(\ref{eq:54}) and (\ref{eq:1}) are decoupled
and the statistical interaction reduces to a (fractional) exclusion condition.

\subsection{Application to quantum 
gas}\label{sec:qgasstatint}   
For a nonrelativistic quantum gas in a box of dimensionality
$\mathcal{D}$ and volume $V=L^\mathcal{D}$ the aforementioned
specifications are encoded in the energy-momentum relation
$\epsilon_0(k)=|\mathbf{k}|^2$ (in units where $\hbar^2/2m=1$) and in a function
$g(\mathbf{k},\mathbf{k}')$. The grand potential (\ref{eq:71})
becomes
\begin{equation}
  \label{eq:12}
  \Omega= -k_BT\left(\frac{L}{2\pi}\right)^\mathcal{D} \int
  d^\mathcal{D}k\,\ln\frac{1+w_\mathbf{k}}{w_{\mathbf{k}}},
\end{equation}
where $w_\mathbf{k}$ is the solution of the nonlinear integral equation
\begin{equation}
  \label{eq:13}
  \frac{|\mathbf{k}|^2-\mu}{k_BT}= \ln(1+w_\mathbf{k}) 
  -\int d^\mathcal{D}k'\,g(\mathbf{k}',\mathbf{k})
\ln\frac{1+w_{\mathbf{k}'}}{w_{\mathbf{k}'}}.
\end{equation}
The particle density in $\mathbf{k}$-space, $\langle n_\mathbf{k}\rangle$, is the
solution of the linear integral equation
\begin{equation}
  \label{eq:14}
  \langle n_{\mathbf{k}}\rangle w_{\mathbf{k}}+
  \int d^\mathcal{D}k'\,g(\mathbf{k},\mathbf{k}')\langle n_{\mathbf{k}'}\rangle=1.
\end{equation}
The fundamental thermodynamic relations (thermodynamic and caloric equations of
state) depend on the solutions of (\ref{eq:13}) and (\ref{eq:14}) as follows:
\begin{equation}
  \label{eq:15}
  \frac{pV}{k_BT} = \left(\frac{L}{2\pi}\right)^\mathcal{D} \int
  d^\mathcal{D}k\,\ln\frac{1+w_\mathbf{k}}{w_{\mathbf{k}}},
\end{equation}
\begin{equation}
  \label{eq:16}
  \mathcal{N}= \left(\frac{L}{2\pi}\right)^\mathcal{D} \int
  d^\mathcal{D}k\, \langle n_{\mathbf{k}}\rangle,
\end{equation}
\begin{equation}
  \label{eq:17}
 U= \left(\frac{L}{2\pi}\right)^\mathcal{D} \int
  d^\mathcal{D}k\, |\mathbf{k}|^2\langle n_{\mathbf{k}}\rangle.
\end{equation}
If the statistical interaction is of the form $g(|\mathbf{k}-\mathbf{k}'|)$ then
the solutions of (\ref{eq:13}) and (\ref{eq:14}) only depend on the magnitude
of the particle momenta.

\subsection{Nonlinear Schr{\"{o}}dinger 
model}\label{sec:apnls}  
Consider the boson gas in $\mathcal{D}=1$ with repulsive contact interaction of
strength $c$ as described by the NLS Hamiltonian
\begin{equation}\label{eq:HbosN}
H = -\sum_{i=1}^N\frac{\partial^2}{\partial x_i^2} +2c\sum_{j<i}\delta(x_i-x_j).
\end{equation}
The thermodynamic Bethe ansatz (TBA) solution \cite{YY69, Yang70, KBI93,
  Taka99} of the NLS model expresses the grand potential in the form
\begin{equation}
  \label{eq:73}
  \Omega(T,L,\mu)=-k_BT\left(\frac{L}{2\pi}\right)\int_{-\infty}^{+\infty}dk
\ln\left(1+e^{-\epsilon(k)/k_BT}\right),
\end{equation}
where $\epsilon(k)$ is the solution of the Yang-Yang equation \cite{YY69},
\begin{equation}
  \label{eq:74}
\epsilon(k)=k^2-\mu-\frac{k_BT}{2\pi}\int_{-\infty}^{+\infty}dk'K(k-k')
\ln\left(1+e^{-\epsilon(k')/k_BT}\right)
\end{equation}
with kernel
\begin{equation}
  \label{eq:3}
  K(k-k')= \frac{2c}{c^2+(k-k')^2}.
\end{equation}
The particle density $\langle n_k\rangle$ is the solution, for given
$\epsilon(k)$, of the Lieb-Liniger equation \cite{LL63, YY69},
\begin{equation}
  \label{eq:4}
 \langle n_k\rangle\left[1+e^{\epsilon(k)/k_BT}\right] = 
1+\frac{1}{2\pi}\int_{-\infty}^{+\infty}dk'\,K(k-k')\langle n_{k'}\rangle.  
\end{equation}
Bernard and Wu \cite{BW94} showed that this TBA solution is equivalent to the
thermodynamics of a statistically interacting gas in $\mathcal{D}=1$ if the
following identifications are made:
\begin{subequations}  
\label{eq:75}
\begin{align}
  w_k & = e^{\epsilon(k)/k_BT}, \\ 
  g(k-k') & =\delta(k-k')-\frac{1}{2\pi}K(k-k').
\end{align}
\end{subequations}

\subsection{Generalization of NLS model}\label{sec:gennls}  
The generalized NLS model is a quantum gas in
$\mathcal{D}$ dimensions with the statistical interaction expressed by the
kernel
\begin{equation}
  \label{eq:64sl}
  K(\mathbf{k}-\mathbf{k}')=
  \frac{2\Gamma(\mathcal{D})}{\pi^{\mathcal{D}/2-1}\Gamma(\mathcal{D}/2)}
  \frac{c^\mathcal{D}}{[c^2+(\mathbf{k}-\mathbf{k}')^2]^\mathcal{D}}
\end{equation}
of the Yang-Yang equation (\ref{eq:74}) and Lieb-Liniger equation (\ref{eq:4})
generalized to $\mathcal{D}\geq1$ and designed to reproduce the exact
thermodynamics of the dynamically interacting NLS model in $\mathcal{D}=1$. The
kernel (\ref{eq:64sl}) has the properties
\begin{subequations}
  \label{eq:65sl}
  \begin{align}
    \lim_{c\to\infty}K(\mathbf{k}) & = 0,
    \\
    \lim_{c\to0}K(\mathbf{k})&=2\pi\delta(\mathbf{k}),
    \\
    \int d^\mathcal{D}k\,K(\mathbf{k})&=2\pi.
  \end{align}
\end{subequations}
This model interpolates between the ideal Fermi-Dirac (FD) gas in the
strong-coupling limit $(c=\infty)$ and the ideal Bose-Einstein (BE) gas in the
weak-coupling limit $(c=0)$ in all dimensions $\mathcal{D}\geq1$. In the limit
$\mathcal{D}\to\infty$ it turns into the ideal FD gas for all $c>0$.

\begin{figure}[htb]
  \centering 
   \includegraphics[width=86mm]{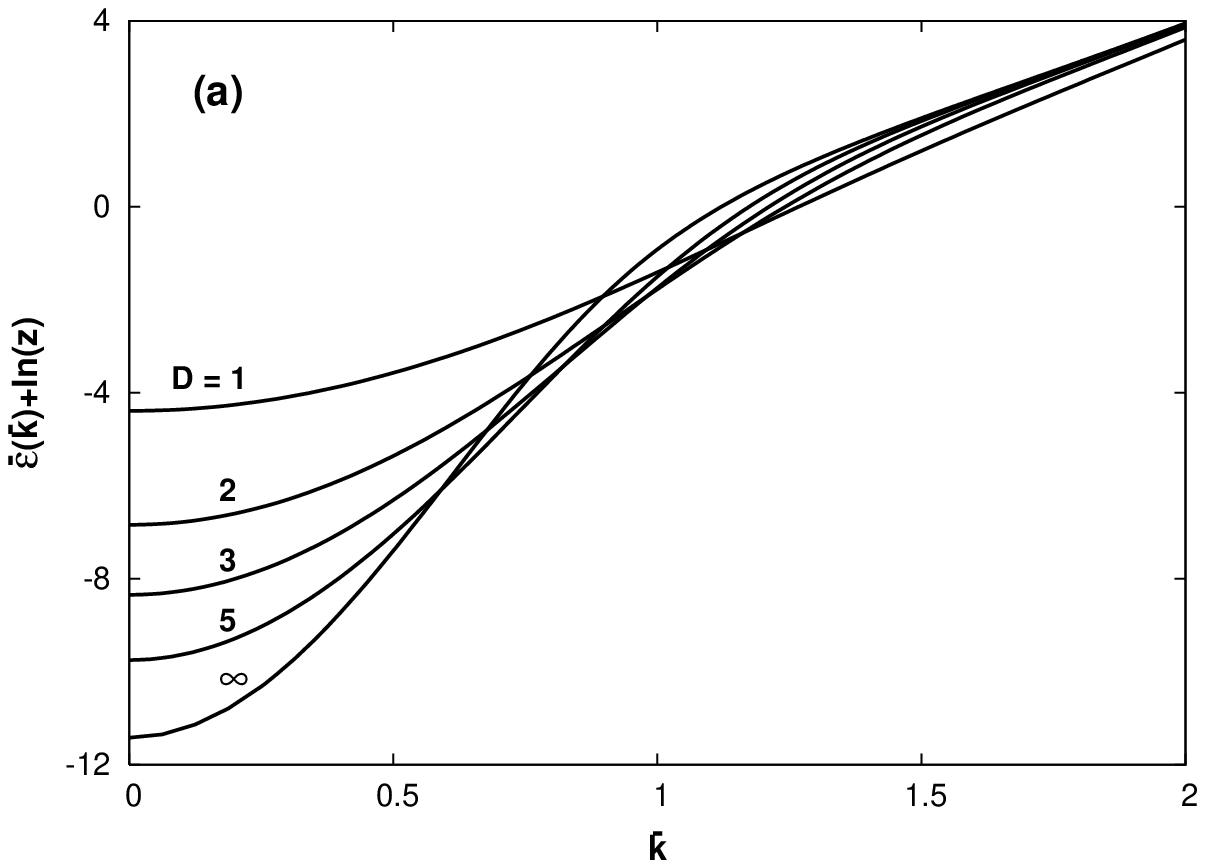}
   \includegraphics[width=86mm]{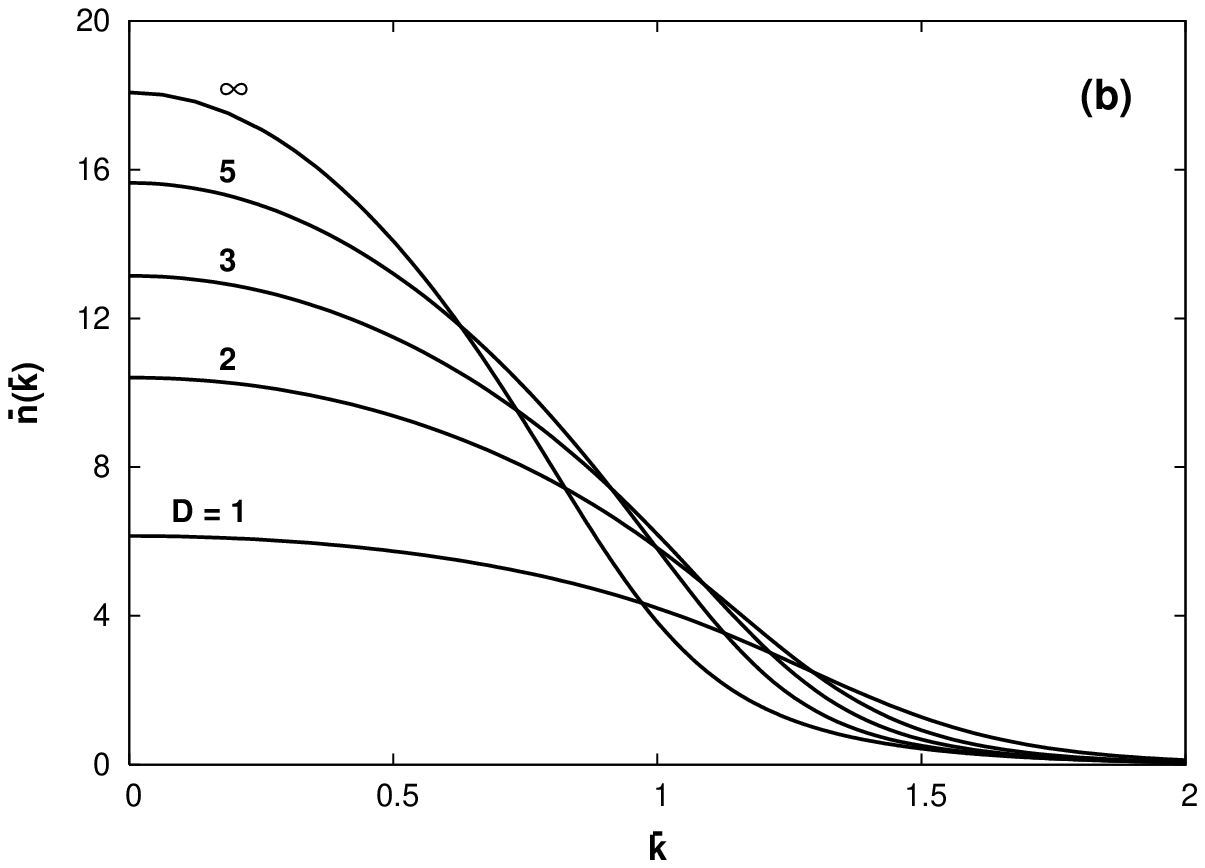}
   \caption{Iterative solutions (a) $\bar{\epsilon}(\bar{k}) +\ln z$ of
     Eq.~(\ref{eq:69dds}), and (b) $\bar{n}(\bar{k})$ of
     Eq.~(\ref{eq:77dds}) for $\bar{c}=0.25$, $z=3$, and various values of
     $\mathcal{D}$. The curves for $\mathcal{D}=\infty$ are solutions
     of Eqs.~(\ref{eq:9}) and (\ref{eq:10}).}
  \label{fig:one}
\end{figure}

For the further analysis of the generalized NLS model we reduce
Eqs.~(\ref{eq:13}) and (\ref{eq:14}) into integral equations for the functions
\begin{equation}
  \label{eq:2}
  \epsilon(k)\doteq k_BT\ln w_\mathbf{k},\quad n(k)\doteq\langle 
n_\mathbf{k}\rangle,
\end{equation}
where $k\doteq|\mathbf{k}|$.  We also introduce scaled quantities
\begin{subequations}
 \label{eq:31tba}
\begin{align}  
  \bar{k}\doteq \frac{k}{\sqrt{k_BT}},\qquad
\bar{c}\doteq \frac{c}{\sqrt{k_BT}}, \\
\bar{\epsilon}(\bar{k})\doteq
  \frac{\epsilon(k)}{k_BT},\qquad     
  \bar{n}(\bar{k})\doteq n(k).
\end{align} 
\end{subequations}
Equations (\ref{eq:74}) and (\ref{eq:4}) thus thermodynamically generalized to
$\mathcal{D}\geq1$ become
\begin{eqnarray}
  \label{eq:69dds}
  \bar{\epsilon}(\bar{k}) = \bar{k}^2-\ln z 
- \int_0^\infty d\bar{k}'\bar{K}(\bar{k},\bar{k}') 
\ln(1+e^{-\bar{\epsilon}(\bar{k}')}),
\end{eqnarray}
\begin{eqnarray}
  \label{eq:77dds}
 \bar{n}(\bar{k})\left[1+e^{\bar{\epsilon}(\bar{k})}\right] = 1 
+ \int_0^\infty d\bar{k}'\bar{K}(\bar{k},\bar{k}')\bar{n}(\bar{k}'),
\end{eqnarray}
with fugacity $z=e^{\mu/k_{B}T}$ and reduced kernel
\begin{eqnarray}
  \label{eq:7}  
   & \nonumber \hspace{-45mm} \bar{K}(\bar{k},\bar{k}') = 
\frac{\displaystyle 2\bar{c}^\mathcal{D}\Gamma(\mathcal{D})}{\displaystyle
  [\Gamma(\mathcal{D}/2)]^2} 
\\ & \times
\frac{\displaystyle \bar{k}'^{\mathcal{D}-1}(\bar{c}^{2}+\bar{k}^{2}+\bar{k}'^{2})
      }{\displaystyle 
[\bar{c}^{4}+2\bar{c}^{2}(\bar{k}^{2}+\bar{k}'^{2})+(\bar{k}^{2}- 
\bar{k}'^{2})^{2}]^{(\mathcal{D}+1)/2}}.
\end{eqnarray}
Any particular solution $\bar{\epsilon}(\bar{k})$, $\bar{n}(\bar{k})$ at fixed $z$,
$\bar{c}$ describes the system over a range of temperature $T$, chemical
potential $\mu$, and coupling constant $c$.  For couplings $0<\bar{c}<\infty$ the
solutions of Eqs.~(\ref{eq:69dds}), (\ref{eq:77dds}) also depend on
$\mathcal{D}$. High-precision data for $\bar{\epsilon}(\bar{k})$, 
$\bar{n}(\bar{k})$
can be obtained from an iterative solution. The sample of data shown in
Fig.~\ref{fig:one} exhibit the main characteristic features of these functions.

The function $\bar{\epsilon}(\bar{k}) +\ln z$ is monotonically increasing from a
minimum at $\bar{k}=0$. The dependence on $z$, $\bar{c}$ (not shown), and
$\mathcal{D}$ is strongest at small $\bar{k}$ and weakens rapidly at
$\bar{k}>1$. The presence of a shoulder combined with a deep minimum at
$\bar{k}=0$ signals a BEC ordering tendency (weakly evident here in the
$\mathcal{D}>2$ data). The function $\bar{n}(\bar{k})$ is monotonically
decreasing from a maximum at $\bar{k}=0$.  Again there is a significant
dependence on $z$, $\bar{c}$, and $\mathcal{D}$ at small $\bar{k}$. The
function itself vanishes rapidly at $\bar{k}>1$.

The dependence on $k$, $\mu$, $T$ of the functions $\bar{\epsilon}(\bar{k})$ and
$\bar{n}(\bar{k})$ is not reducible, for $0<\bar{c}<\infty$, to a dependence on a
single variable, $(k^2-\mu)/k_BT$. This represents a major deviation from a
signature property of ideal gases.

\subsection{Limiting cases}\label{sec:limcas}  
In the strong-coupling and weak-coupling limits, which are fixed points of the
scaling relation for the coupling constant, we recover the familiar FD and BE
results:
\begin{equation}  
  \label{eq:38tba} 
\bar{\epsilon}(\bar{k})=\bar{k}^2-\ln z,\quad \bar{n}(\bar{k})=  
\left(\frac{1}{z}e^{\bar{k}^2}+1\right)^{-1},   
\end{equation}  
for $\bar{c}=\infty$, and 
\begin{equation}  
  \label{eq:39tba}  
\bar{\epsilon}(\bar{k})=\ln\left(\frac{1}{z}e^{\bar{k}^2}-1\right),\quad  
\bar{n}(\bar{k})= \left(\frac{1}{z}e^{\bar{k}^2}-1\right)^{-1}
\end{equation}  
for $\bar{c}=0$.  The $\mathcal{D}$-independence of these functions
is another signature property of ideal gases, a property upheld in the
presence of fractional statistics \cite{PMK07}.  The parabolic curve,
$\bar{\epsilon}(\bar{k})+\ln z=\bar{k}^2$, may serve as baseline for the
solutions of (\ref{eq:69dds}). It is exact in the strong-coupling limit for
all $z>0$, but only for $z\to0$ in the weak-coupling limit.  For $0<z<1$ the
deviations of $\bar{\epsilon}(\bar{k})+\ln z$ from $\bar{k}^2$ and of
$\bar{n}(\bar{k})$ from zero are suppressed by factors $\sim e^{-\bar{k}^2}$
as $\bar{k}$ increases.  We have noted very similar behavior in the numerical
results for $0<c<\infty$.

In the limit $\mathcal{D}\to\infty$ the reduced kernel (\ref{eq:7})
acquires a much simpler structure,
\begin{equation}
  \label{eq:8}
  \lim_{\mathcal{D}\to\infty}\bar{K}(\bar{k},\bar{k}')
  =\delta\left(\bar{k}'-\sqrt{\bar{k}^2+\bar{c}^2}\right), 
\end{equation}
but the $\bar{c}$-dependence is retained and (for $\bar{c}>0$) also the
statistical coupling of particles with distinct momenta. Equations
(\ref{eq:69dds}) and (\ref{eq:77dds}), with reduced kernel (\ref{eq:8}) turn
into the implicit functions
\begin{equation}
  \label{eq:9}
  \bar{\epsilon}(\bar{k}) = \bar{k}^2-\ln{z} 
  -\ln\left(1+e^{-\bar{\epsilon}\left(\sqrt{\bar{k}^2+\bar{c}^2}\right)}\right),
\end{equation}
\begin{equation}
  \label{eq:10}
   \bar{n}(\bar{k})\left[1+e^{\bar{\epsilon}(\bar{k})}\right] =
   1+\bar{n}\left(\sqrt{\bar{k}^2+\bar{c}^2}\right). 
\end{equation}
The solution of Eq.~\eqref{eq:9} is
\begin{align}
  \label{eq:26}
  e^{-\bar{\epsilon}(\bar{k})} = 
    \sum_{l=0}^{\infty}z^{l+1}e^{-(l+1)\bar{k}^{2}}e^{-c^{2}l(l+1)/2},
\end{align}
and the solution of Eq.~(\ref{eq:10}) is
\begin{align}
  \label{eq:30}
  \bar{n}(\bar{k}) = \sum_{l=0}^{\infty}\alpha_{l}z^{l+1}e^{-\bar{k}^{2}(l+1)}
\end{align}
with the recursion
\begin{align}
  \label{eq:31}\nonumber
  \alpha_{l} = e^{-\frac{c^{2}}{2}l(l+1)} &
  -\sum_{j=1}^{l}\left( \alpha_{l-j}e^{-\frac{c^{2}}{2}j(j-1)} 
\right. \nonumber \\
& \left.- \alpha_{j-1}e^{-\frac{c^{2}}{2}((l-j)(l-j+1)+2j)} 
  \right).
\end{align}
For $\bar{c}\to\infty$ Eqs.~(\ref{eq:38tba}) are recovered, and for
$\bar{c}\to0$ Eqs.~(\ref{eq:39tba}).

%
\section{Thermodynamic analysis of 
generalized NLS model}\label{sec:thenls}    
%
Exact results for the thermodynamics of the generalized NLS model in
$\mathcal{D}\geq1$ and across the range $0\leq c\leq\infty$ of coupling
strengths can now be calculated from the solutions $\bar{\epsilon}(\bar{k})$
and $\bar{n}(\bar{k})$ of Eqs.~(\ref{eq:69dds}) and (\ref{eq:77dds}),
respectively.

\subsection{NLS functions}\label{sec:nlsfunc}  
The fundamental thermodynamic relations (\ref{eq:15})--(\ref{eq:17}) are 
rewritten in the form
\begin{equation}
  \label{eq:2sl2}
  \frac{p\lambda_T^\mathcal{D}}{k_BT}= F_p^{(\mathcal{D})}(z,\bar{c}),
\end{equation}
\begin{equation}
  \label{eq:3sl2}
  \frac{\mathcal{N}\lambda_T^\mathcal{D}}{V}=
  F_\mathcal{N}^{(\mathcal{D})}(z,\bar{c})\quad \left[+\frac{z}{1-z}\right],
\end{equation}
\begin{equation}
  \label{eq:4sl2}
   \frac{U\lambda_T^\mathcal{D}}{V}\left\slash \frac{\mathcal{D}}{2}k_BT\right.= 
 F_U^{(\mathcal{D})}(z,\bar{c}),
\end{equation}
where
\begin{equation}
  \label{eq:11}
  \lambda_T\doteq \sqrt{\frac{h^2}{2\pi mk_BT}} 
~\stackrel{\hbar^2/2m=1}{\longrightarrow}~
  \sqrt{\frac{4\pi}{k_BT}} 
\end{equation}
is the thermal wavelength and the term in (\ref{eq:3sl2}) enclosed by
square-brackets is relevant only if $\bar{c}=0$ and $\mathcal{D}>2$. The NLS
functions in (\ref{eq:2sl2})-(\ref{eq:4sl2}) are defined as follows:
\begin{equation}
  \label{eq:18}
  F_p^{(\mathcal{D})}(z,\bar{c}) \doteq\frac{2}{\Gamma(\mathcal{D}/2)}\int_0^\infty
  d\bar{k}\,\bar{k}^{\mathcal{D}-1} \ln\left(1+e^{-\bar{\epsilon}(\bar{k})}\right),
\end{equation}
\begin{equation}
  \label{eq:19}
F_\mathcal{N}^{(\mathcal{D})}(z,\bar{c}) \doteq\frac{2}{\Gamma(\mathcal{D}/2)}
\int_0^\infty d\bar{k}\,\bar{k}^{\mathcal{D}-1}\bar{n}(\bar{k}),
\end{equation}
\begin{equation}
  \label{eq:20}
F_U^{(\mathcal{D})}(z,\bar{c}) \doteq\frac{2}{\Gamma(\mathcal{D}/2+1)}
\int_0^\infty d\bar{k}\,\bar{k}^{\mathcal{D}+1}\bar{n}(\bar{k}),
\end{equation}
where the additional dependence of $\bar{\epsilon}(\bar{k})$, $\bar{n}(\bar{k})$ on
$\mathcal{D}, \bar{c}, z$ is implied.

In the strong-coupling and weak-coupling limits, the NLS functions turn into
the familiar FD functions,
\begin{equation}
  \label{eq:25fd}
  f_{n}(z)\doteq
  \frac{1}{\Gamma(n)}\int_{0}^{\infty}
\frac{dx\,x^{n-1}}{z^{-1}e^{x}+1},\quad z\geq 0,
\end{equation}
and BE functions, 
\begin{equation}
  \label{eq:32be}
  g_{n}(z)\doteq
  \frac{1}{\Gamma(n)}\int_{0}^{\infty}
\frac{dx\,x^{n-1}}{z^{-1}e^{x}-1},\quad 0\leq z\leq 1,
\end{equation}
respectively: 
\begin{subequations}
  \label{eq:5} 
\begin{align}
  F_p^{(\mathcal{D})}(z,\infty)  & =   F_U^{(\mathcal{D})}(z,\infty) =
  f_{\mathcal{D}/2+1}(z), \\ 
  F_\mathcal{N}^{(\mathcal{D})}(z,\infty) & = f_{\mathcal{D}/2}(z),
\end{align}
\end{subequations}
and
\begin{subequations}
  \label{eq:6} 
\begin{align}
   F_p^{(\mathcal{D})}(z,0) &= F_U^{(\mathcal{D})}(z,0) =
   g_{\mathcal{D}/2+1}(z),    \\
   F_\mathcal{N}^{(\mathcal{D})}(z,0) &= g_{\mathcal{D}/2}(z).
 \end{align}
\end{subequations}
Furthermore, for $\mathcal{D}\gg1$ fermionic behavior results for any
$\bar{c}>0$:
\begin{subequations}
  \label{eq:21}
\begin{align}
  F_p^{(\mathcal{D})}(z,\bar{c}),~
  F_U^{(\mathcal{D})}(z,\bar{c}) &\stackrel{\mathcal{D}\gg1}{\leadsto}
  f_{\mathcal{D}/2+1}(z),  \\
F_\mathcal{N}^{(\mathcal{D})}(z,\bar{c}) 
&\stackrel{\mathcal{D}\gg1}{\leadsto}
  f_{\mathcal{D}/2}(z).
\end{align}
\end{subequations}
With increasing $\mathcal{D}$ the factor $\bar{k}^{\mathcal{D}\pm 1}$ pushes
all significant contributions to the integrals (\ref{eq:18})-(\ref{eq:20})
toward larger and larger $\bar{k}$, where the deviations of
$\bar{\epsilon}(\bar{k})$ and $\bar{n}(\bar{k})$ from their
$(\bar{c}=\infty)$-values (\ref{eq:38tba}) become smaller and smaller
\cite{fn4}.

A characteristic ideal-gas property is that the dependence of the fugacity on
the thermodynamic variables $T,V,\mathcal{N}$ is expressible as a function of a
single variable,
\begin{equation}
  \label{eq:24}
  x\doteq\lambda_Tv^{-1/ \mathcal{D}},\quad v\doteq V/ \mathcal{N}. 
\end{equation}
In the Maxwell-Boltzmann (MB) gas we have $x^\mathcal{D}=z$, in the FD gas
$x^\mathcal{D}=f_{\mathcal{D}/2}(z)$, and in the BE gas
$x^\mathcal{D}=g_{\mathcal{D}/2}(z)$. A unique functional relation persists in
the case of fractional statistics \cite{PMK07}. In the generalized NLS model,
however, we have $x^\mathcal{D}=F_\mathcal{N}^{(\mathcal{D})}(z,\bar{c})$ with
a separate $T$-dependence contained in $\bar{c}$.  For ideal quantum gases,
including those with fractional statistics, there also exists a unique
functional dependence of $pV/ \mathcal{N}k_BT$ on $z$. Again this no longer
holds in the generalized NLS model, where we have $pV/ \mathcal{N}k_BT=
F_p^{(\mathcal{D})}(z,\bar{c}) /F_\mathcal{N}^{(\mathcal{D})}(z,\bar{c})$.

\subsection{Reference values}\label{sec:refval}  
We introduce reference values for the thermodynamic variables $v$, $T$, $p$
based on the thermal wavelength $\lambda_T$ and the MB equation of state
$pv=k_BT$ in the presentation of our data below:
\begin{equation}
  \label{eq:108sl2}
  k_BT_v = \frac{4\pi}{v^{2/ \mathcal{D}}},\quad p_v = \frac{4\pi}{v^{2/
        \mathcal{D}+1}}\qquad (v = \mathrm{const.}) 
\end{equation}
\begin{equation}
  \label{eq:109sl2}
  v_T = \left(\frac{4\pi}{k_BT}\right)^{\mathcal{D}/2},\quad
  p_T = \frac{(k_BT)^{\mathcal{D}/2+1}}{(4\pi)^{\mathcal{D}/2}}\qquad 
(T = \mathrm{const.})
\end{equation}
\begin{equation}
  \label{eq:110sl2}
  k_BT_p = 4\pi\left(\frac{p}{4\pi}\right)^{\frac{2}{\mathcal{D}+2}},\quad  
v_p = \left(\frac{4\pi}{p}\right)^{\frac{\mathcal{D}}{\mathcal{D}+2}}\qquad 
(p = \mathrm{const.})
\end{equation}
They are especially useful in comparative plots that encompass the full
range of $\bar{c}$ at finite $\mathcal{D}$.

For the thermodynamic analysis we must adapt the NLS functions to the type of
process under consideration. Each function has a different $z$-dependence at
fixed $c$, depending, for example, on whether we consider $v=\mathrm{const}$,
$T=\mathrm{const}$, or $p=\mathrm{const}$. To this end we introduce three kinds
of reduced coupling constants for use in isochoric, isothermal, and
isobaric processes, respectively:
    \begin{equation}
    \label{eq:7sl2}
    c_v\doteq\frac{c}{\sqrt{k_BT_v}}= 
\bar{c}\sqrt{\frac{T}{T_v}}=\frac{\bar{c}}{x}\qquad (v=\mathrm{const}),
  \end{equation}
\begin{equation}
    \label{eq:6sl2}
    c_T\doteq\frac{c}{\sqrt{k_BT}} =\bar{c}\qquad (T=\mathrm{const}),
  \end{equation}
 \begin{equation}
    \label{eq:9sl2}
    c_p\doteq\frac{c}{\sqrt{k_BT_p}}= 
\bar{c}\sqrt{\frac{T}{T_p}}\doteq\frac{\bar{c}}{y}\qquad (p=\mathrm{const}),
  \end{equation}
where $x$ and $y$ are the solutions of 
  \begin{equation}
    \label{eq:8sl2}
    x^\mathcal{D}=F_\mathcal{N}^{(\mathcal{D})}(z,c_vx),
  \end{equation}
  \begin{equation}
    \label{eq:10sl2}
        y^{\mathcal{D}+2}=F_p^{(\mathcal{D})}(z,c_py),
  \end{equation}
respectively.

Reference values based on the chemical potential present themselves as an
alternative in some situations. Defining $\ln z\doteq\bar{T}_v/T$ in isochoric
processes and $\ln z\doteq\bar{T}_p/T$ in isobaric processes, we have
\begin{equation}
  \label{eq:22}
\hspace*{-15mm}  \frac{\bar{T}_v}{T_v}=\frac{\bar{p}_v}{p_v}
=\left[\Gamma\left(\frac{\mathcal{D}}{2}+1\right)\right]^{\frac{2}{\mathcal{D}}}
~\stackrel{\mathcal{D}\gg1}{\leadsto}~ \frac{\mathcal{D}}{2e}
\end{equation}
for $v=\mathrm{const}$, and  
\begin{equation}
  \label{eq:23}
\hspace*{-15mm}   \frac{\bar{T}_p}{T_p}=\frac{\bar{v}_p}{v_p}
=\left[\Gamma\left(\frac{\mathcal{D}}{2}+2\right)\right]^{\frac{2}{\mathcal{D}+2}}
~\stackrel{\mathcal{D}\gg1}{\leadsto}~
\left(\frac{\mathcal{D}}{2}+1\right)\frac{1}{e}
\end{equation}
for $p=\mathrm{const}$.  The divergence of these ratios of in the limit
$\mathcal{D}\to\infty$ has some surprising consequence as will be discussed in
Sec.~\ref{sec:pddi}.

%
\section{Results}\label{sec:selres}  
%

In Ref.~\cite{PMK07} we presented a panoramic view of the thermodynamics
of the generalized CS model (ideal quantum gas with fractional statistics) in
$\mathcal{D}$ dimensions. The emphasis was on the crossover between boson-like
and fermion-like features in isochores, isotherms, isobars, response functions,
and the speed of sound as caused by aspects of the statistical interaction that
reflect long-range attraction and short-range repulsion.

The generalized NLS model considered here exhibits some similarities with the
generalized CS model regarding thermodynamic properties, especially their
dependence on the coupling constants of the two models. However, there are
notable differences, many of which can be identified as significant deviations
from ideal gas behavior. In our presentation of results we highlight these
deviations and the role of dimensionality.

\subsection{Isochores, isobars, and isotherms}
\label{sec:isoch}  
The dependences of $p$ on $T$ at $v=\mathrm{const}$, of $v$ on $T$ at
$p=\mathrm{const}$, and of $p$ on $v$ at $T=\mathrm{const}$ are determined by
(\ref{eq:2sl2}) and (\ref{eq:3sl2}) in parametric representations,
\begin{equation}
  \label{eq:27}
    \frac{p}{p_v}=
  \frac{F_p^{(\mathcal{D})}(z,c_vx)}
{[F_\mathcal{N}^{(\mathcal{D})}(z,c_vx)]^{1+\frac{2}{\mathcal{D}}}},\quad 
\frac{T}{T_v}=
\left[F_\mathcal{N}^{(\mathcal{D})}(z,c_vx)\right]^{\frac{-2}{\mathcal{D}}},
\end{equation}
\begin{equation}
  \label{eq:28}
   \frac{v}{v_p}=
  \frac{[F_p^{(\mathcal{D})}(z,c_py)]^{\frac{\mathcal{D}}{\mathcal{D}+2}}}
{F_\mathcal{N}^{(\mathcal{D})}(z,c_py)},\quad 
\frac{T}{T_p}=\left[F_p^{(\mathcal{D})}(z,c_py)\right]^{\frac{-2}{\mathcal{D}+2}},
\end{equation}
\begin{equation}
  \label{eq:29}
   \frac{p}{p_T}=
  F_p^{(\mathcal{D})}(z,c_T),\quad 
\frac{v}{v_T}=\left[F_\mathcal{N}^{(\mathcal{D})}(z,c_T)\right]^{-1}, 
\end{equation}
respectively, with the fugacity $z$ in the role of the parameter. Here $x$ and
$y$ are the solutions of (\ref{eq:8sl2}) and (\ref{eq:10sl2}), respectively.

\begin{figure}[htb]
  \centering
   \includegraphics[width=86mm]{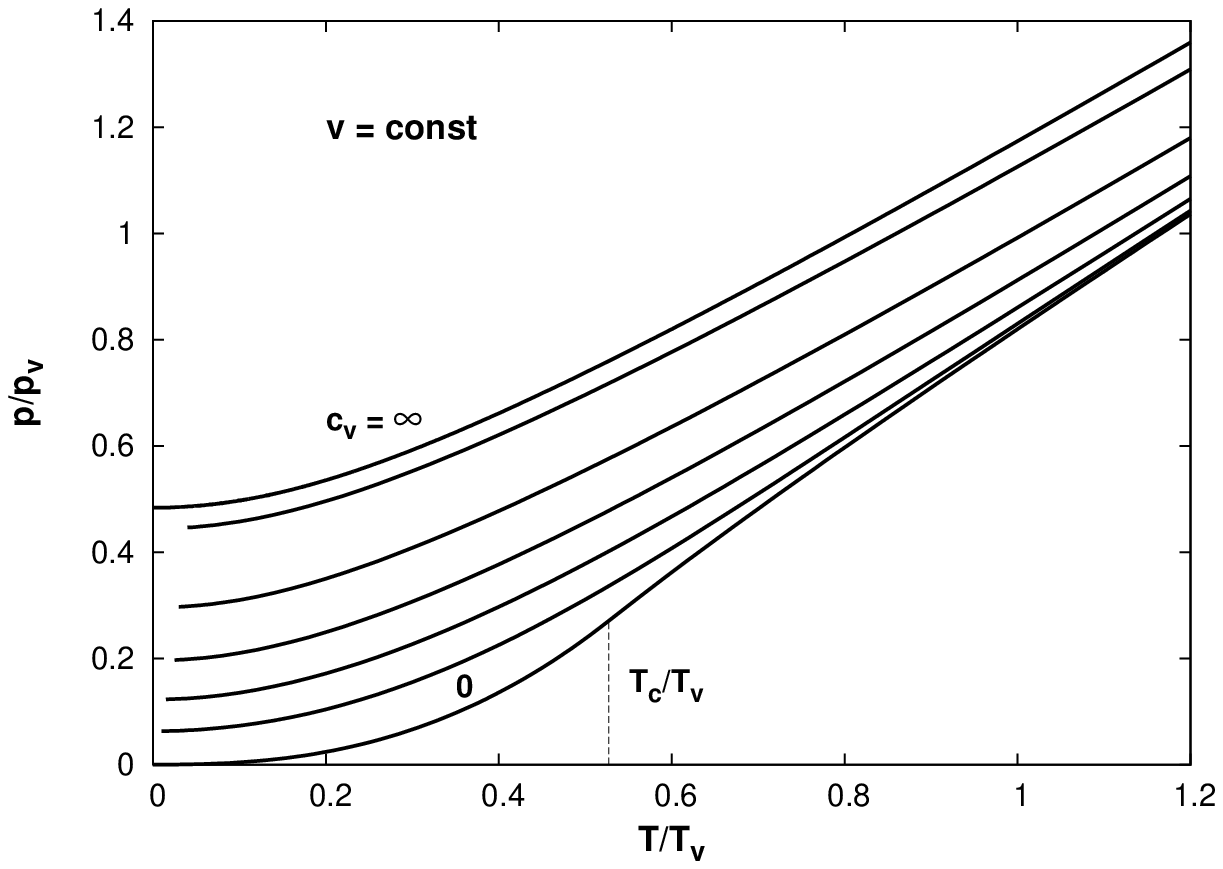}
   \includegraphics[width=86mm]{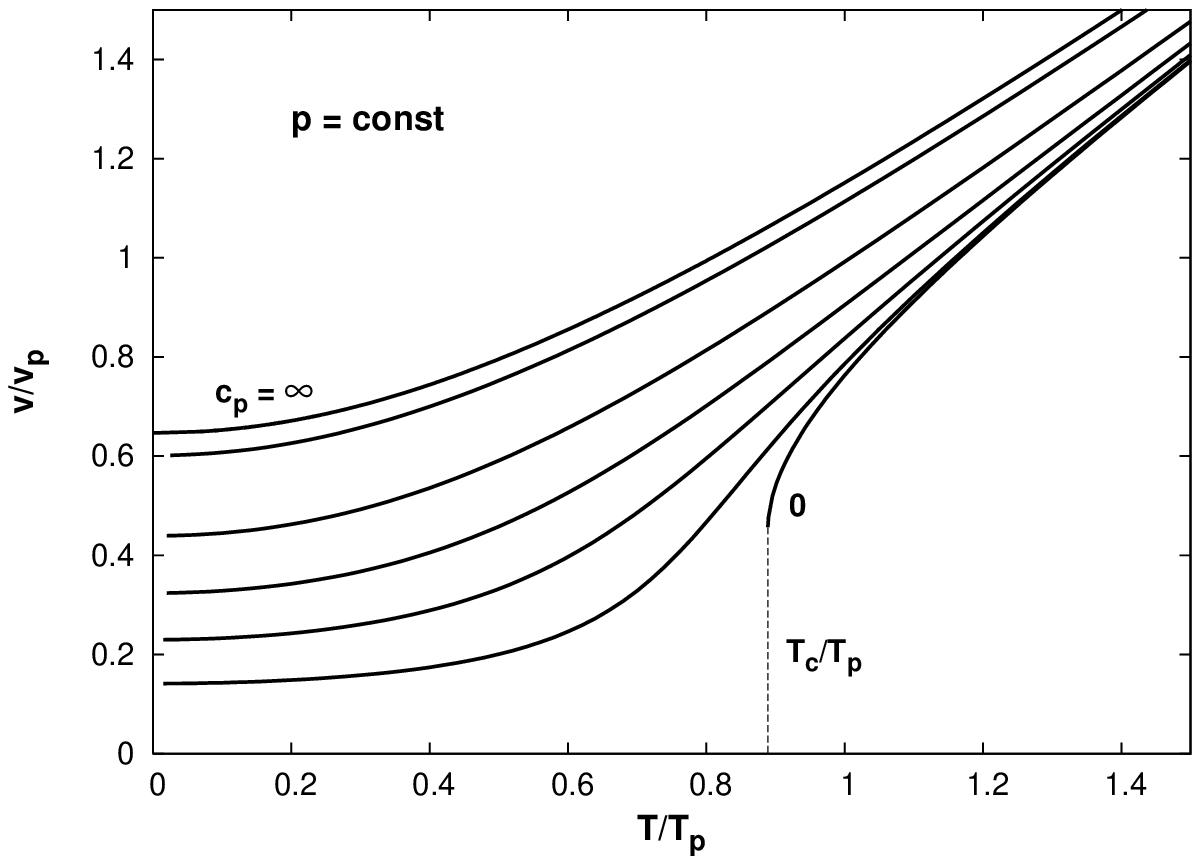}
   \hspace*{3mm}\includegraphics[width=84mm]{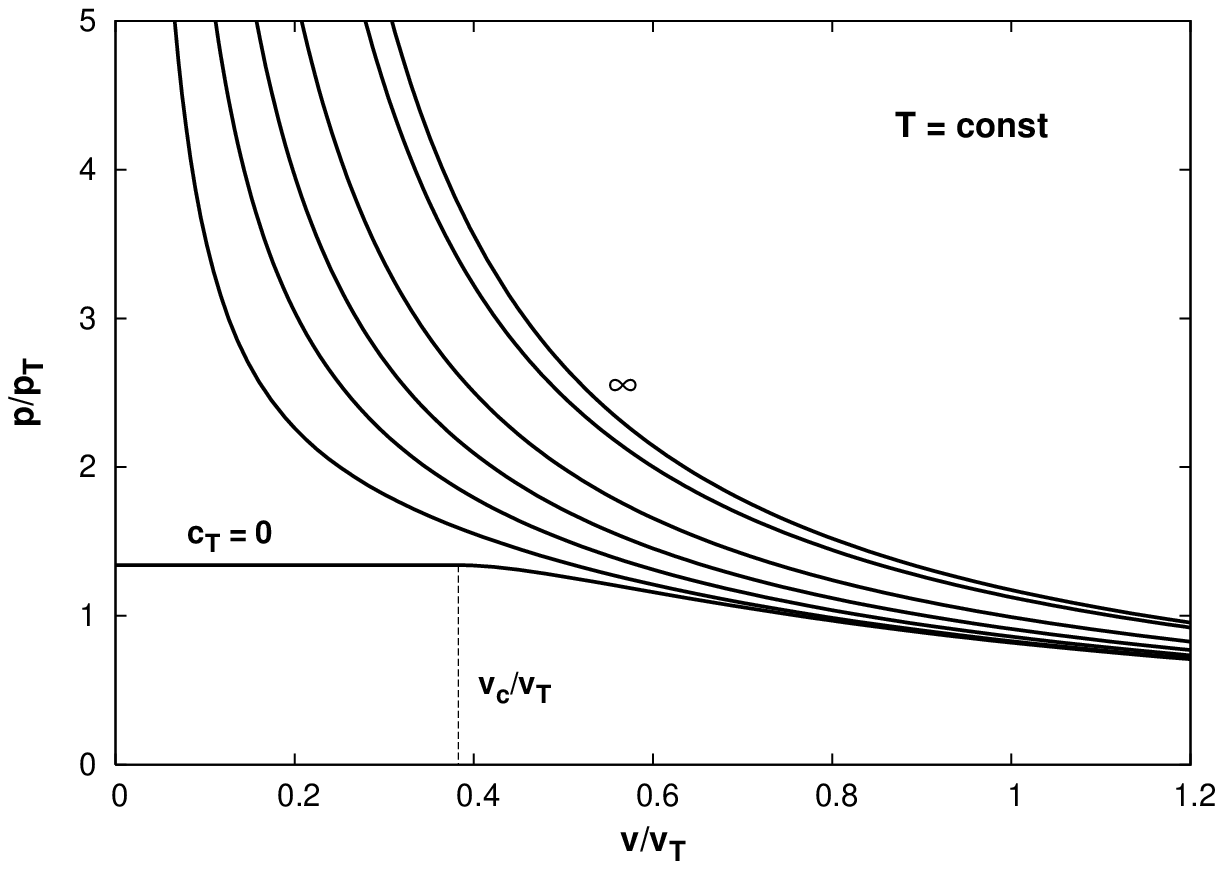}
\caption{Isochores, isobars, and isotherms in $D = 3$ for $c_{v,p,T}=0$, $0.1$,
  $0.25$, $0.5$, $1$, $3$, $\infty$ (from bottom up in each frame). }
  \label{fig:two}
\end{figure}

In Fig.~\ref{fig:two} we show isochores, isobars, and isotherms for various
coupling strengths $c_{v,p,T}$ in $\mathcal{D}=3$. The variation of the curves
between the (weak-coupling) boson limit and the (strong-coupling) fermion limit
is similar to what was observed in the generalized CS model \cite{PMK07}: the
convergence of all curves toward the MB line at high $T$ or large $v$, and the
fanning out at low $T$ or small $v$. Corresponding plots in other $\mathcal{D}$
show similar trends in the two models.

The shape of the curves for $c_{v,p,T}>0$ in Fig.~\ref{fig:two} yield some
insight into the physical interpretation of the statistical interaction. For
weak couplings $(c_{v,p,T}<1)$ the curves exhibit boson-like features at high
$T$ or large $v$ and fermion-like features at low $T$ or small $v$. These
observations translate into a long-range attractive part and a shorter-range
repulsive part of the statistical interaction. The attractive tail is only
present for small $c_{v,p,T}$, whereas the repulsive core is conspicuous for all
$c_{v,p,T}>0$. 

Among all the curves only the ones pertaining to the boson limit
$(c_{v,p,T}=0)$ have a singularity. This singularity signals the presence of a
phase transition, the onset of BEC. In the $\mathcal{D}=3$ case shown, the
phase transition occurs at $T_c/T_v\simeq0.527$, $p_c/p_v\simeq0.271$ along the
isochore, at $T_c/T_p\simeq0.889$, $v_c/v_p\simeq0.456$ along the isobar, and
at $v_c/v_T\simeq0.383$, $p_c/p_T\simeq1.341$ along the isotherm.

The bosonic isochore has a singularity at $T_c/T_v>0$ only in $\mathcal{D}>2$.
In $2<\mathcal{D}\leq4$ it has a discontinuity in curvature. In $\mathcal{D}>4$ it
becomes a discontinuity in slope. In the limit $\mathcal{D}\to\infty$ the bosonic
isochore itself becomes discontinuous.  By contrast, the bosonic isobar has a
singularity at $T_c/T_p>0$ in all dimensions $\mathcal{D}\geq1$, but with
$v_c/v_p>0$ only in $\mathcal{D}>2$.  The bosonic isotherm has a horizontal
portion at $v/v_T<v_c/v_T$ in $\mathcal{D}>2$ (see Ref.~\cite{PMK07} for more
details on the bosonic curves.)

We have already noted that all three NLS functions (\ref{eq:18})-(\ref{eq:20})
converge toward the corresponding FD functions as $\mathcal{D}\to\infty$
provided we have $\bar{c}>0$. One reflection of this fact in the data for
isochores, isobars, and isotherms is that all curves for $c_{v,p,T}>0$ move
closer together as $\mathcal{D}$ increases. They coalesce into the universal
curve (isochore, isobar, or isotherm) representing the ideal FD gas in
$\mathcal{D}=\infty$. Only the bosonic curves at $T<T_{c}$ or $v<v_{c}$ stay
apart.

\begin{figure}[htb]
  \centering
   \hspace{-3mm}\includegraphics[width=89mm]{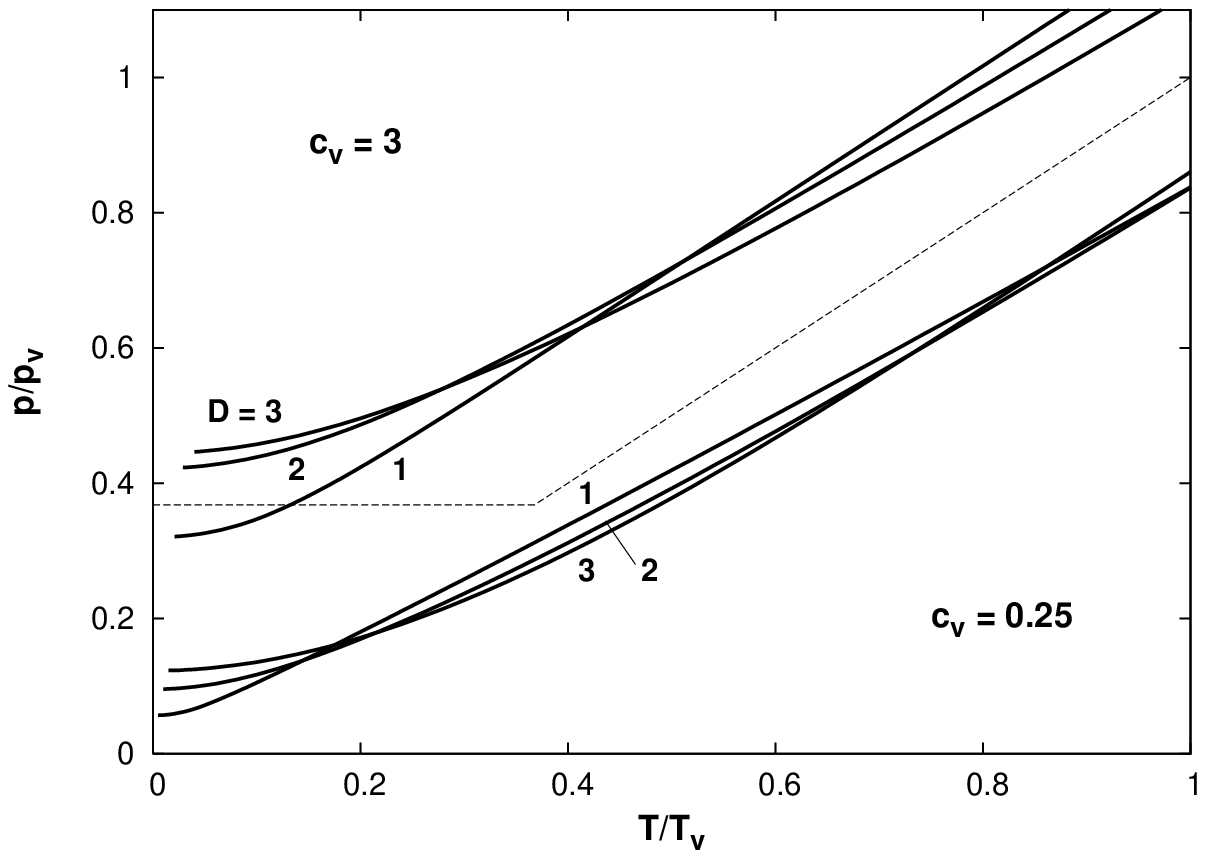}
   \includegraphics[width=86mm]{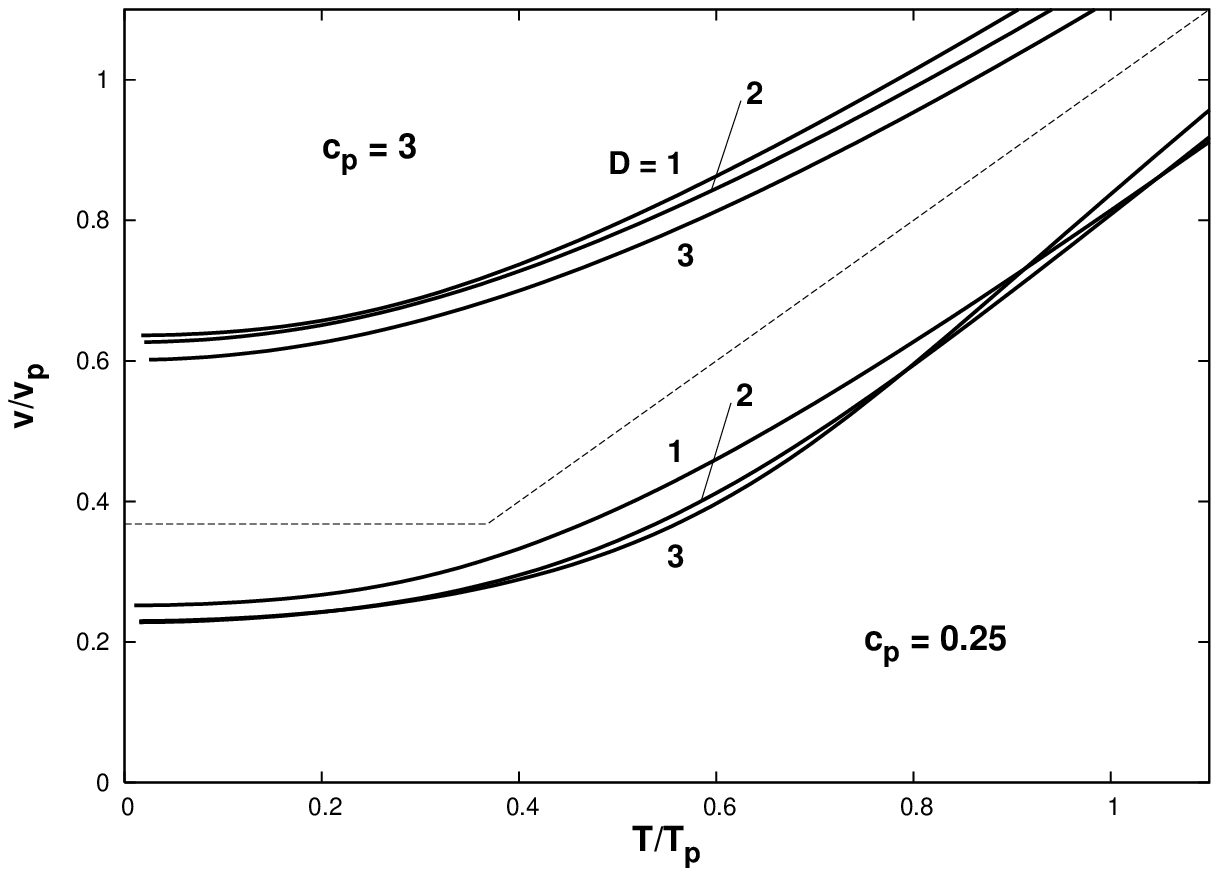}
   \includegraphics[width=86mm]{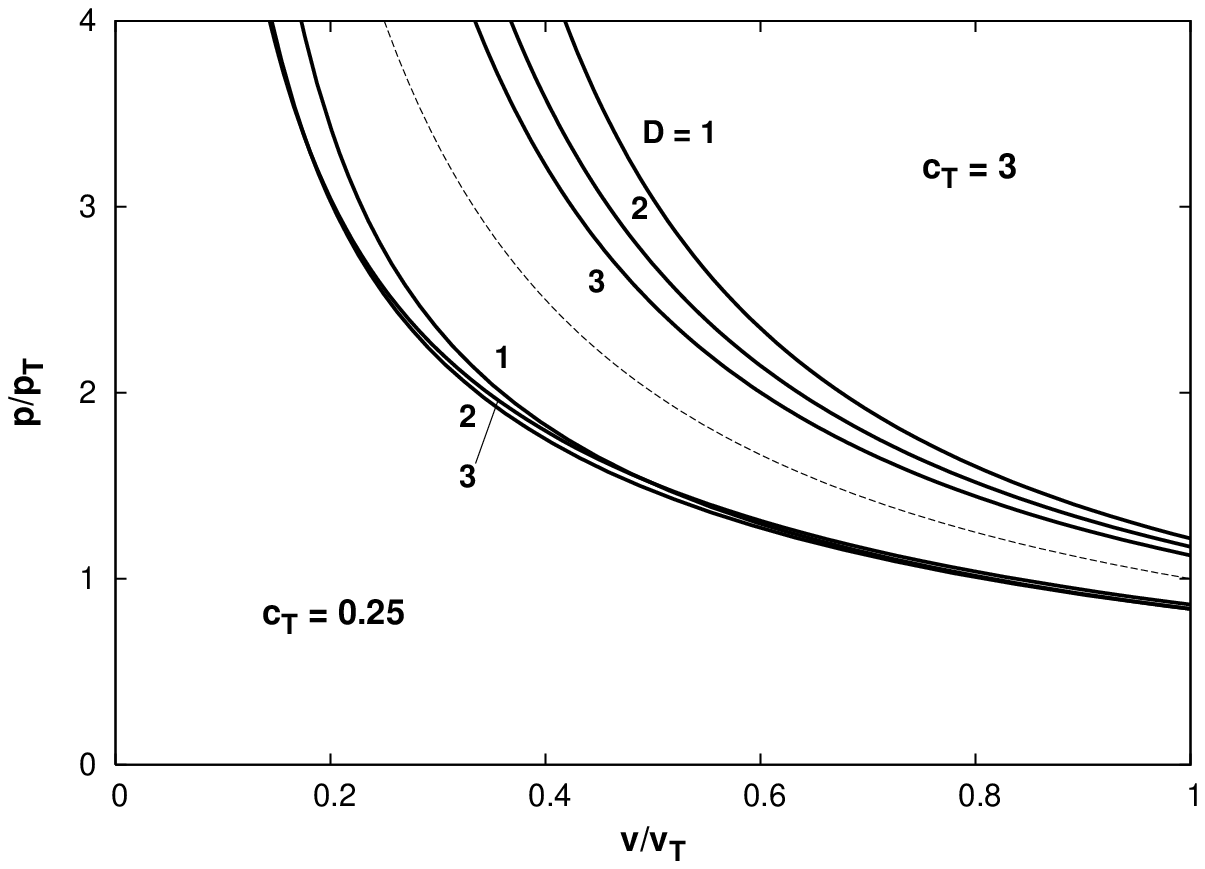}
\caption{Isochores, isobars, and isotherms in $D = 1,2,3$ for $c_{v,p,T}=0.25$
  and $c_{v,p,T}=3$. The dashed lines represent the curves for
  $\mathcal{D}=\infty$.}
  \label{fig:four}
\end{figure}

In Fig.~\ref{fig:four} we show two sets of isochores, isobars, and isotherms
for the generalized NLS model in $\mathcal{D}=1,2,3$, one set for weak
coupling, the other for strong coupling. Also shown (dashed) are the
corresponding curves pertaining to $\mathcal{D}=\infty$, which will be derived
in Sec.~\ref{sec:pddi}. For the most part, the weak-coupling and
strong-coupling curves are located on opposite sides of the dashed line in each
frame.

Convergence of the data for $\mathcal{D}=1,2,3$ toward the line representing
$\mathcal{D}=\infty$ is only apparent at high $T$ or large $v$ and more clearly in
the strong-coupling data than in the weak-coupling data. This is not surprising
in view of the observation made earlier in the context of Fig.~\ref{fig:two}
that for weak couplings the (effectively) long-range attractive part and
short-range repulsive part of the statistical interaction are responsible for
opposing trends and a crossover between them. However, convergence becomes
manifest in higher $\mathcal{D}$ (not shown) as the NLS functions gradually turn
into FD functions first for strong couplings and then also for weak couplings.

\subsection{Phase transition in 
$\mathcal{D}=\infty$}\label{sec:pddi}  

It is well known that no phase transition at $T>0$ exists for free fermions in
$\mathcal{D}<\infty$. No transition is expected to exist in the generalized
NLS model in finite $\mathcal{D}$ except in the boson limit. However, a curious
transition does emerge in the limit $\mathcal{D}\to\infty$, where the
generalized NLS model with $\bar{c}>0$ effectively turns into an ideal FD gas.

\begin{figure}[htb]
  \centering
  \includegraphics[width=86mm]{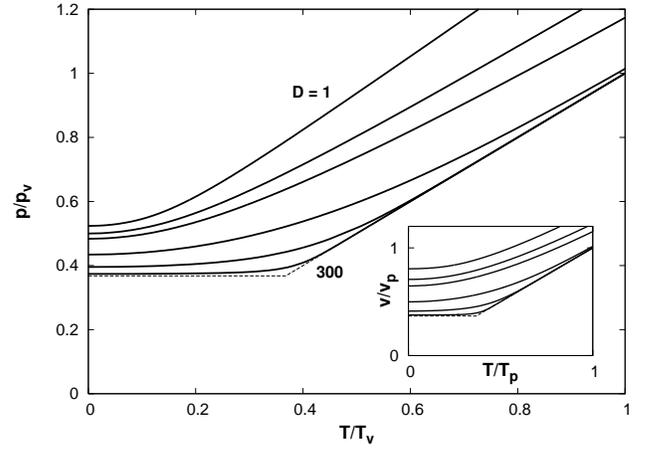}
  \caption{Isochores (main plot) and isobars (inset) of the ideal FD gas in
    $\mathcal{D}=1$, $2$, $3$, $10$, $40$, $300$ (top to bottom). The
    result for $\mathcal{D}=\infty$ is represented by the dashed lines.}
  \label{fig:three}
\end{figure}

To determine the thermodynamic equation of state, $f(p,v,T)=0$, of the
generalized NLS model in $\mathcal{D}=\infty$ we recall (\ref{eq:21}) and
rewrite (\ref{eq:27})-(\ref{eq:29}) for $\mathcal{D}\gg1$ with the FD functions
substituted for the NLS functions. A singularity at $T>0$ results as a
consequence of the fact that the two limits $\mathcal{D}\to\infty$,
$z\to\infty$ are not interchangeable.  The emergence of the singularity is
apparent in the isochores and isobars in $\mathcal{D}\gg1$ as shown in
Fig.~\ref{fig:three}.

The isochore at $z<\infty$ in the limit $\mathcal{D}\to\infty$ yields a
straight-line segment with unit slope and zero intercept in the
$(T/T_{v},p/p_{v})$-plane over a nonzero interval $T_{c}\leq T\leq \infty$:
\begin{equation}
  \label{eq:42}
 \frac{p}{p_v} ~\stackrel{\mathcal{D}\gg1}{\simeq}~ \frac{T}{T_v}
 \frac{f_{\mathcal{D}/2+1}(z)}{f_{\mathcal{D}/2}(z)}
 ~\stackrel{\mathcal{D}\to\infty}{\longrightarrow}~ \frac{T}{T_{v}}.
\end{equation}
The reference values (\ref{eq:108sl2}) become $k_{B}T_{v}=4\pi$, $p_{v}=4\pi/v$
in the limit $\mathcal{D}\to\infty$. The same isochore in the limit
$\mathcal{D}\to\infty$, $z\to\infty$ with $\mathcal{D}/2=r\ln z$, $r\geq 0$
yields a horizontal line segment over a nonzero interval $0\leq T\leq T_{c}$:
\begin{equation}
  \label{eq:43}
 \frac{p}{p_v} ~\stackrel{\mathcal{D}\gg1}{\simeq}~ \frac{f_{\mathcal{D}/2+1}(z)}
{[f_{\mathcal{D}/2}(z)]^{1+2/ \mathcal{D}}}  ~\stackrel{z\gg1}{\leadsto}~
\frac{e^{-1}}{1+2/ \mathcal{D}}, 
\end{equation}
\begin{equation}
  \label{eq:44}
  \frac{T}{T_v}  ~\stackrel{\mathcal{D}\gg1}{\simeq}~
\left[f_{\mathcal{D}/2}(z)\right]^{-2/ \mathcal{D}} ~\stackrel{z\gg1}{\leadsto}~ 
\frac{\mathcal{D}}{2}\frac{e^{-1}}{\ln z}, 
\end{equation}
where we have used the leading term in the asymptotic expansion of the FD
function \cite{Path72}. The value of $T_{c}$ is determined by the intersection
point of the two line segments. The equation of state thus reads \cite{fn2}
\begin{equation}
  \label{eq:45}
  pv=\left\{ 
\begin{array}{ll}
k_{B}T, & T>T_{c} \\
k_{B}T_{c}, & T<T_{c} 
\end{array} \right.;\qquad k_{B}T_{c}=\frac{4\pi}{e}.
\end{equation}
This same universal relation can also be inferred from Eq.~(\ref{eq:28}) for
the isobar or from Eq.~(\ref{eq:29}) for the isotherm by performing the
appropriate limits.
All isotherms are hyperbolas, including
the transition line at $T=T_{c}$. All the isochores and isobars consist of two
straight-line segments with the singularity at $T=T_{c}$ as already shown in
Fig.~\ref{fig:three}. 

This somewhat unusual phase transition from a fully intact Fermi sea at
$T<T_{c}$ to an ideal MB gas at $T>T_{c}$ results from the conspiracy of two
opposing effects, one suppressing thermal excitations at low $T$ and the other
enhancing them at high $T$. Both effects grow stronger in higher dimensions.

We know from (\ref{eq:22}) that as $\mathcal{D}$ increases the reference
temperature $T_v$ becomes smaller and smaller compared to the Fermi temperature
$\bar{T}_v$ in isochoric processes (considered here for specificity).  This
suppresses any rise in pressure at sufficiently small but nonzero $T/T_v$ more
and more strongly. In the limit $\mathcal{D}\to\infty$, as $T_v/
\bar{T}_v\to0$, the pressure will remain constant over a non-vanishing interval
of $T/T_{v}$ at the value $p/p_v=e^{-1}$ exerted by the perfect Fermi sea.

We also know (e.g. from analogies to microcanonical ensembles) that as
$\mathcal{D}$ increases the energy density of one-particle states is
progressively thinned out inside the surface of the Fermi hypersphere except
close to the surface. The consequence is that a smaller and smaller amount of
thermal energy is needed to knock out the vast majority of particles from the
Fermi sea. Moreover, the density of vacancies near the Fermi edge becomes so
large that the constraint on occupancy imposed by the Pauli principle is
negligible.

In $\mathcal{D}\gg1$, therefore, if $T/T_v$ is raised gradually, no significant
thermal excitations take place initially because $T_v/ \bar{T}_v\ll1$. The
isochore stays flat. Once $T$ has reached a certain threshold the Fermi sea is
emptied quickly because of its shallowness and the abundance of vacancies close
by. The system thus crosses over from a near perfectly degenerate Fermi sea to
a nearly ideal MB gas on a very short interval of $T/T_{v}$ as documented in
Fig.~\ref{fig:three}. In $\mathcal{D}=\infty$ this crossover has sharpened into a
phase transition. There is no latent heat involved in that transition and there
is no sudden increase in pressure \cite{fn5}.

Note that on the alternative temperature
scale $\bar{T}_{v}$ the emergent crossover between near
perfect Fermi sea and almost ideal MB gas is pushed to lower and lower values of
$T$ as $\mathcal{D}$ increases, ultimately to $T/ \bar{T}_{v}=0$ for
$\mathcal{D}=\infty$. The resultant isochore is then that of the ideal
MB gas all the way down.

It is interesting to recall the phase diagram of the ideal BE gas in
$\mathcal{D}=\infty$ for comparison. The thermodynamic equation of state inferred
from the scaled isochores, isobars, or isotherms as derived, for example, in
Ref.~\cite{PMK07} has the form
\begin{equation}
  \label{eq:42a}
   pv=\left\{ 
\begin{array}{ll}
k_{B}T, & T>T_{c} \\
0, & T<T_{c} 
\end{array} \right.;\qquad k_{B}T_{c}=4\pi.
\end{equation}
As in the FD case there are two phases separated by a transition line at
constant $T$. The high-$T$ phase is again an ideal MB gas. The low-$T$ phase is
a pure BEC. The transition is of first order and occurs at a higher
temperature than in the FD case. Whereas the FD transition disappears in
$\mathcal{D}<\infty$, the BE transition persists down to $\mathcal{D}>2$, but is of
second-order in $\mathcal{D}<\infty$ and occurs along a line in $(p,v,T)$-space
that is no longer an isotherm.

\subsection{Response functions}\label{sec:refunc}  
The three major response functions for a gas of spinless particles are the
isochoric heat capacity, $C_v\doteq\mathcal{N}^{-1}(\partial U/ \partial T)_v$,
the isobaric expansivity, $\alpha_p\doteq v^{-1}(\partial v/ \partial T)_p$,
and the isothermal compressibility, $\kappa_T\doteq -v^{-1}(\partial v/
\partial p)_T$. For the generalized NLS model we must evaluate the expressions
\begin{equation}
  \label{eq:34}
   \frac{C_v}{k_B} = \left(\frac{\mathcal{D}^2}{4} +\frac{\mathcal{D}}{2}\right)
  \frac{F_U^{(\mathcal{D})}(z,c_vx)}{F_\mathcal{N}^{(\mathcal{D})}(z,c_vx)} 
- \frac{\mathcal{D}^2}{4}
  \frac{\frac{\partial}{\partial z}F_U^{(\mathcal{D})}(z,c_vx)}
{\frac{\partial}{\partial
      z}F_\mathcal{N}^{(\mathcal{D})}(z,c_vx)},
\end{equation}
\begin{equation}
  \label{eq:35}
  T_p\alpha_p = \frac{T_p}{T}\left[\left(\frac{\mathcal{D}}{2} +1\right)
  \frac{F_p^{(\mathcal{D})}(z,c_py)\frac{\partial}{\partial z}
F_\mathcal{N}^{(\mathcal{D})}(z,c_py)}
  {F_\mathcal{N}^{(\mathcal{D})}(z,c_py)\frac{\partial}{\partial z}
F_p^{(\mathcal{D})}(z,c_py)} - 
  \frac{\mathcal{D}}{2}\right],
\end{equation}
\begin{equation}
  \label{eq:36}
  p_T\kappa_T = \frac{v}{v_T}\frac{\frac{\partial}{\partial z}
F_{\mathcal{N}}^{(\mathcal{D})}(z,c_T)}
  {\frac{\partial}{\partial z}F_p^{(\mathcal{D})}(z,c_T)},
\end{equation}
versus the independent variables $T/T_v$, $T/T_p$, $v/v_T$, respectively, from
(\ref{eq:27})-(\ref{eq:29}).

\begin{figure}[htb]
  \centering
   \includegraphics[width=86mm]{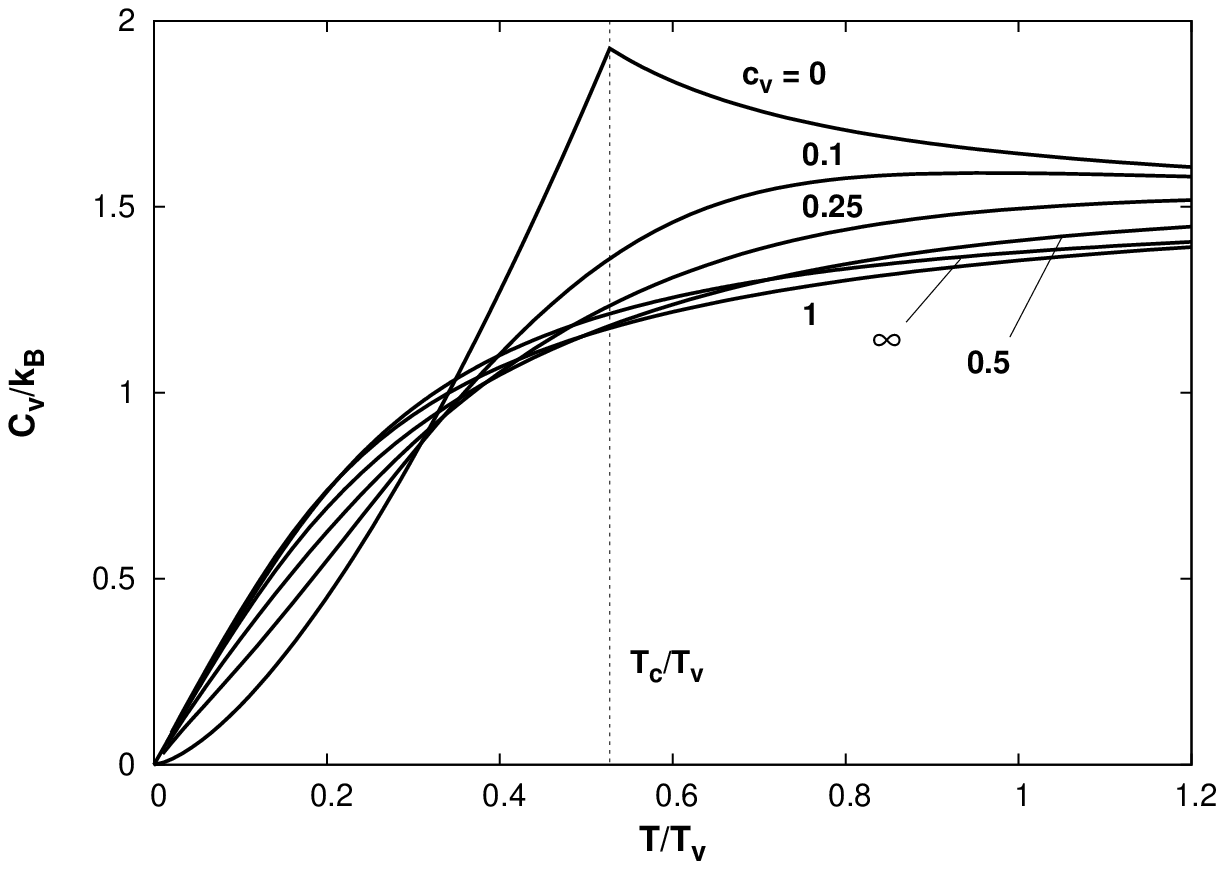}
   \hspace*{2mm}\includegraphics[width=83mm]{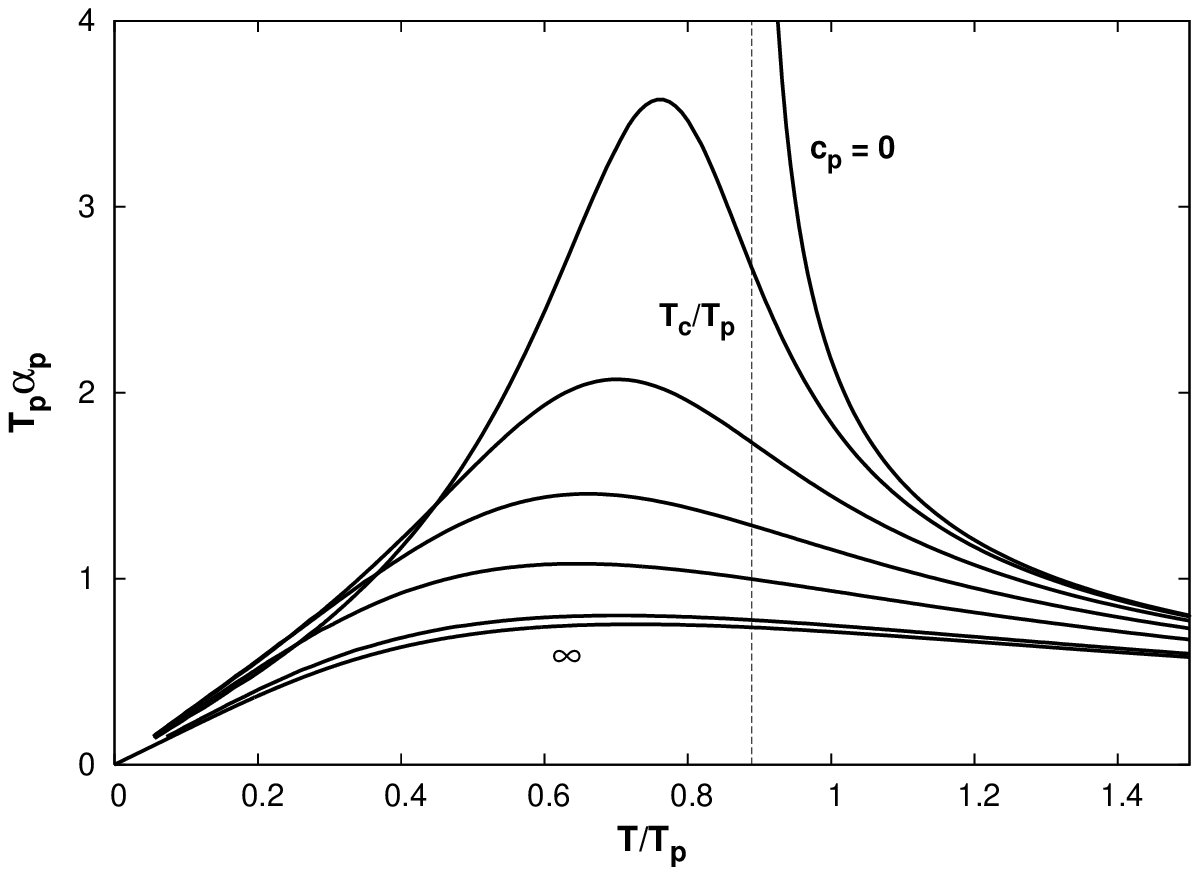}
   \includegraphics[width=86mm]{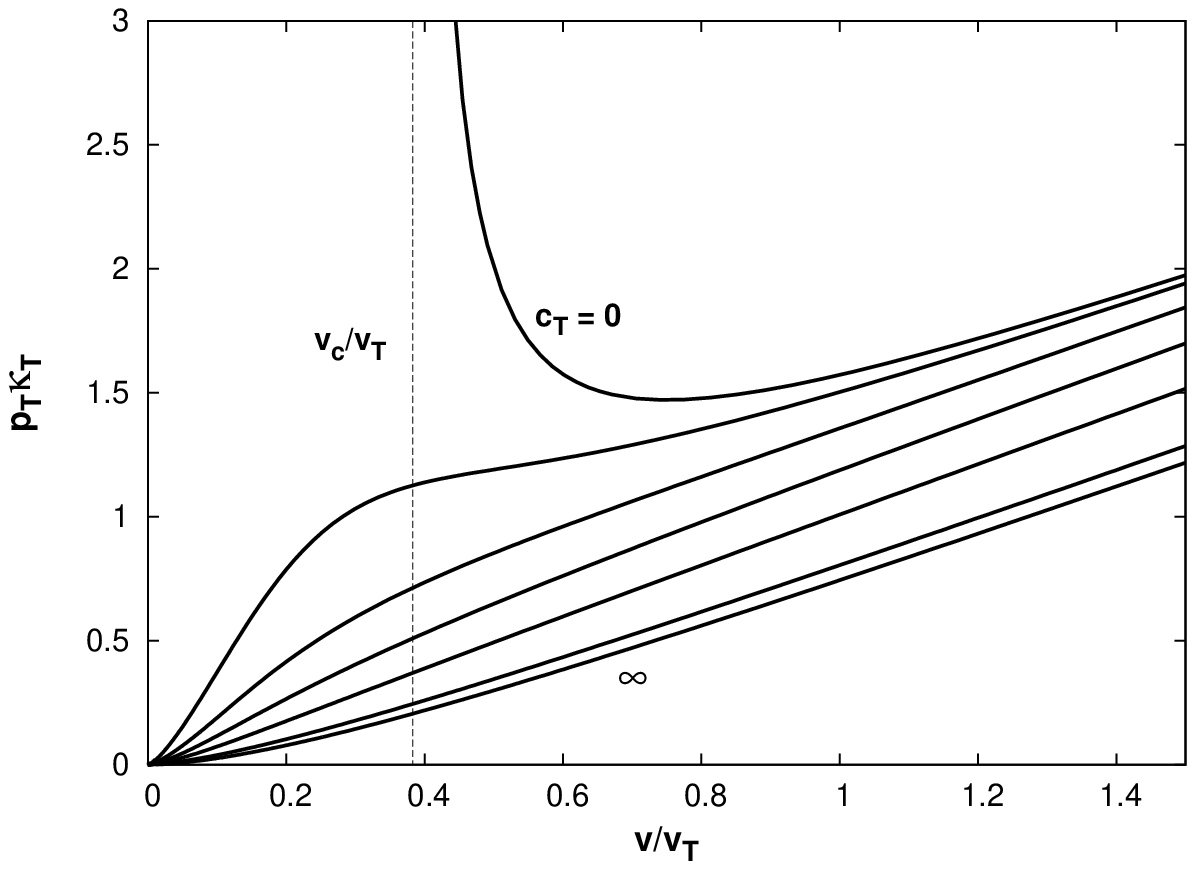}
   \caption{Isochoric heat capacity, isobaric expansivity, and isothermal
     compressibility in $D = 3$ for $c_{v,p,T}=0$, $0.1$, $0.25$, $0.5$, $1$,
     $3$, $\infty$. The heat-capacity curves for $c_v=3$, $\infty$ are
     unresolved. The dashed lines mark the onset of BEC.}
  \label{fig:five}
\end{figure}

In Fig.~\ref{fig:five} we show the dependence of each response function on the
coupling strength in $\mathcal{D}=3$. The variation of the curves between the
BE and FD limits shows some resemblance to that observed in an ideal gas with
fractional statistics (generalized CS model) \cite{PMK07}.  All three response
functions depend only weakly on the statistical interaction at high temperature
or low density. The dominant trends there reflect MB behavior,
$C_{v}=(3/2)k_{B}T$, $\alpha_{p}=1/T$, $\kappa_{T}=v/k_{B}T$. Distinct
boson-like and fermion-like features and crossovers between them emerge at low
temperatures and high densities. The exact analytic behavior of the response
functions in any $\mathcal{D}$ for the FD and BE limits was described in
Ref.~\cite{PMK07}.

The heat capacity in $\mathcal{D}=3$ for strong coupling is dominated by
fermion-like features at all $T$, exhibiting a monotonic descent from the MB
asymptote as $T$ is lowered and a linear approach to zero. For weak coupling
the initial increase, the smooth maximum followed by a steep descent is a
boson-like feature. The ultimate linear approach to zero signals the crossover
to fermion-like behavior.

The expansivity in $\mathcal{D}=3$ depends only weakly on $T$ for strong
coupling and approaches zero linearly as $T\to0$, which is a characteristic
fermion-like behavior. For weak coupling the pronounced rise in expansivity is
a boson-like feature. However, the repulsive core of the statistical
interaction for $\bar{c}>0$, no matter how weak, prevents the expansivity from
diverging and forces the fermion-like behavior at low $T$.

Stiff resistance to compression, perhaps the most outstanding fermion-like
feature, makes itself manifest with growing strength in the strong-coupling
compressibility curves in $\mathcal{D}=3$ as the density is increased. In the
weak-coupling curves, on the other hand, we observe trends reminiscent of
bosonic behavior at moderate densities. While the BE curve diverges at
$v=v_{c}$, the repulsive core of the statistical interaction prevents the
transition from taking place if $\bar{c}>0$. The compressibility curve bends
into a smooth maximum or a mere shoulder down to fermionic stiffness.

\begin{figure}[tb]
  \centering                            
   \includegraphics[width=86mm]{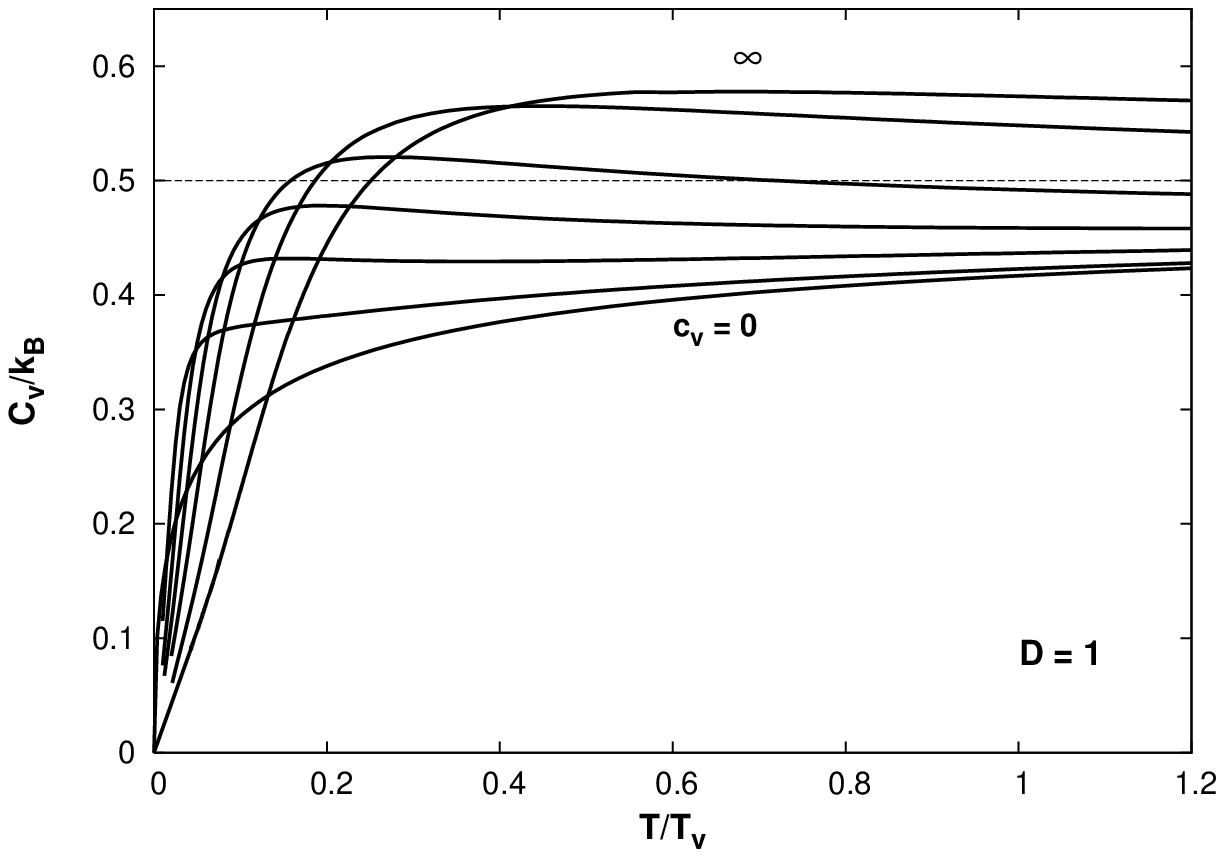}
   \includegraphics[width=86mm]{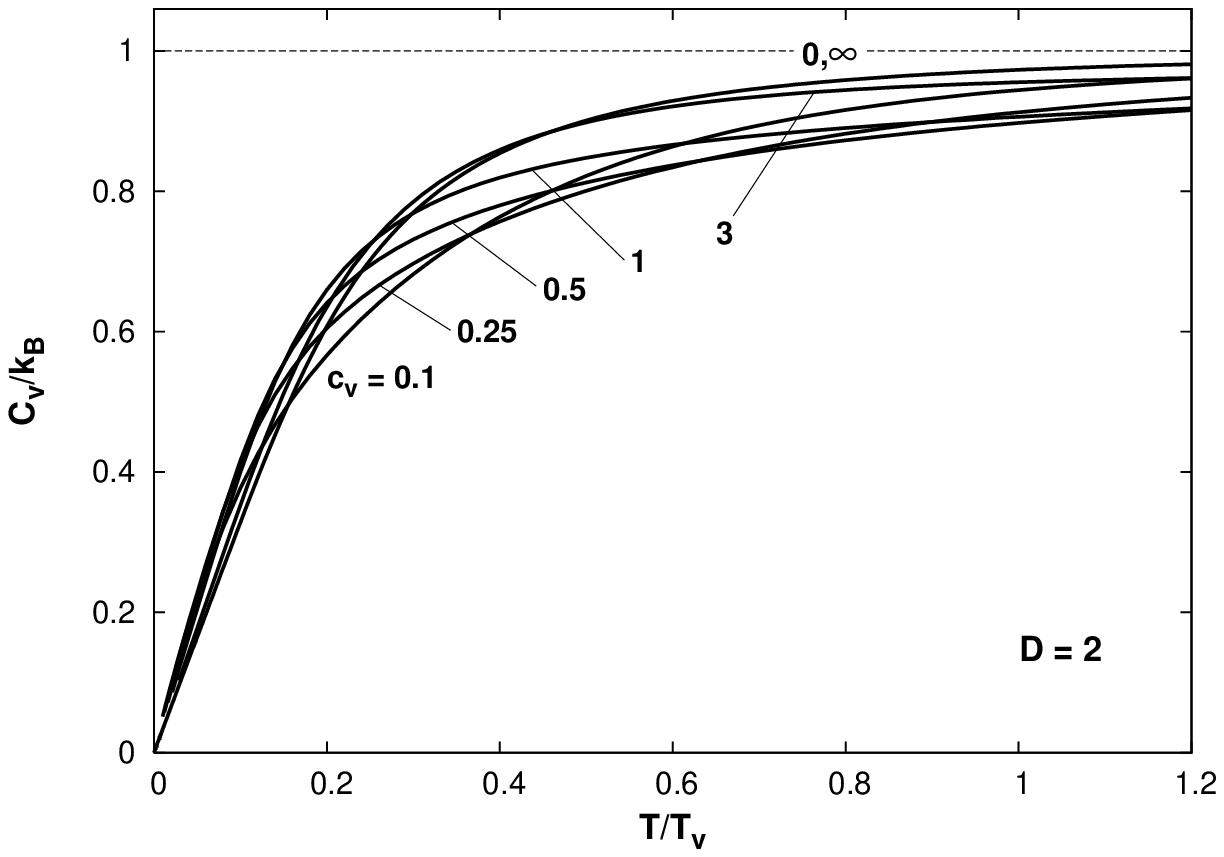}
 \caption{Isochoric heat capacity in $D = 1,2$ for 
   $c_v=0$, $0.1$, $0.25$, $0.5$, $1$, $3$, $\infty$. The dashed lines reflect the
   high-$T$ asymptotic values $C_{v}/k_{B}=\mathcal{D}/2$.}
  \label{fig:six}
\end{figure}

Similarities to the response functions of an ideal quantum gas with fractional
statistics are also manifest in other dimensions $\mathcal{D}$. However, there
are two notable exceptions.  In Fig.~\ref{fig:six} we show the heat capacity for
$\mathcal{D}=1,2$ in the same format as the data for $\mathcal{D}=3$ in
Fig.~\ref{fig:five}.

In $\mathcal{D}=2$ the isochoric heat capacity of an ideal quantum gas is
well-known not to depend on the exclusion statistics \cite{May64, VRH95,
  Iguc97c, Aoya01, Lee97, SC04, Angh02}. That is no longer the case in the
presence of a statistical interaction such as realized in the generalized NLS
model. Only the two curves representing the weak-coupling and strong-coupling
limits coincide. The curves at intermediate coupling are subject to shifting
trends caused by the long-range attractive and short-range repulsive parts of
the statistical interaction.

In $\mathcal{D}=1$ the heat capacity curves are monotonically increasing
functions near the BE limit and functions with one smooth maximum near the FD
limit.  Upon variation of the exclusion statistical parameter between the two
limits in the quantum ideal gas the appearance of the smooth maximum coincides
with a switch in sign of the leading correction to the high-$T$ asymptote
\cite{PMK07}.  Upon variation of the statistical coupling strength between the
same limits in the generalized NLS model, the smooth maximum at low $T$ appears
before the approach to the asymptote switches side from below the asymptote to
above it. In consequence there is a range of coupling strengths where the heat
capacity has a smooth maximum followed by a smooth minimum.

\begin{figure}[htb]
  \centering
   \includegraphics[width=86mm]{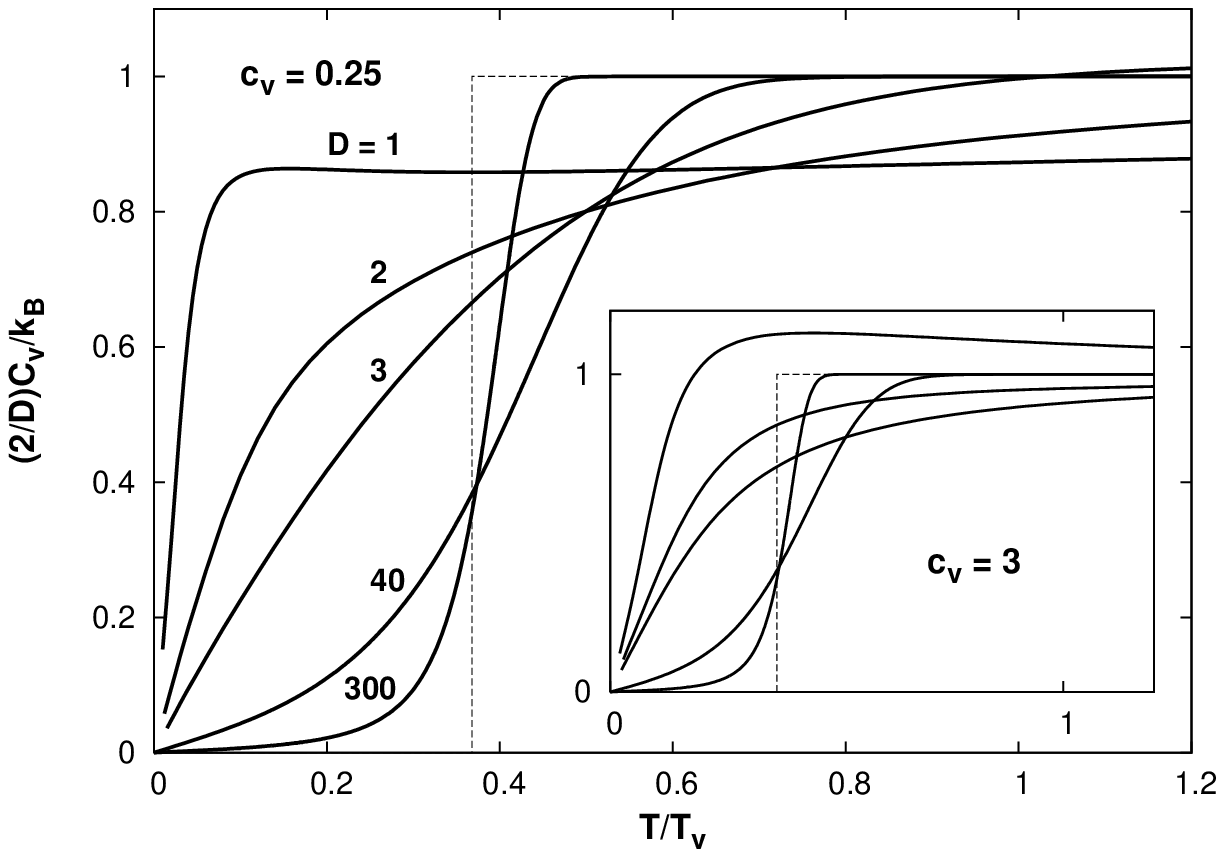}
   \includegraphics[width=86mm]{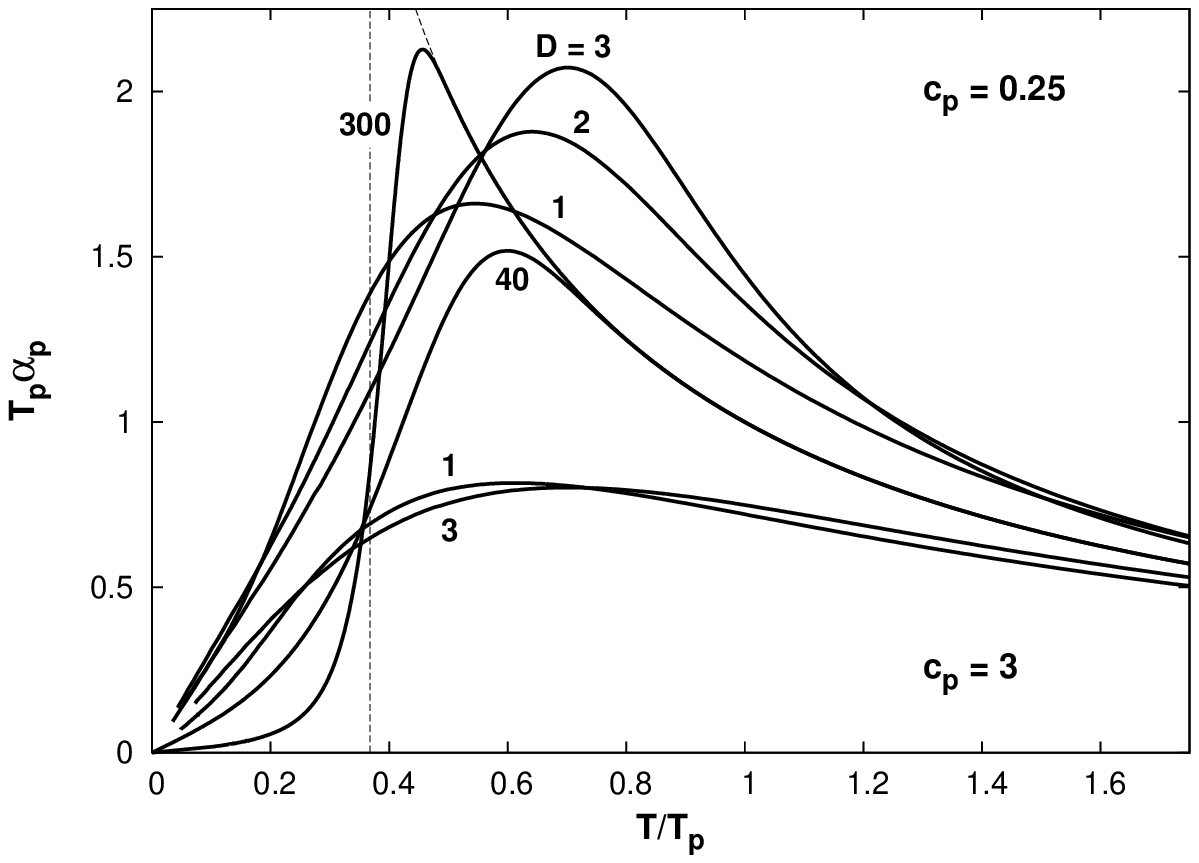}
   \includegraphics[width=86mm]{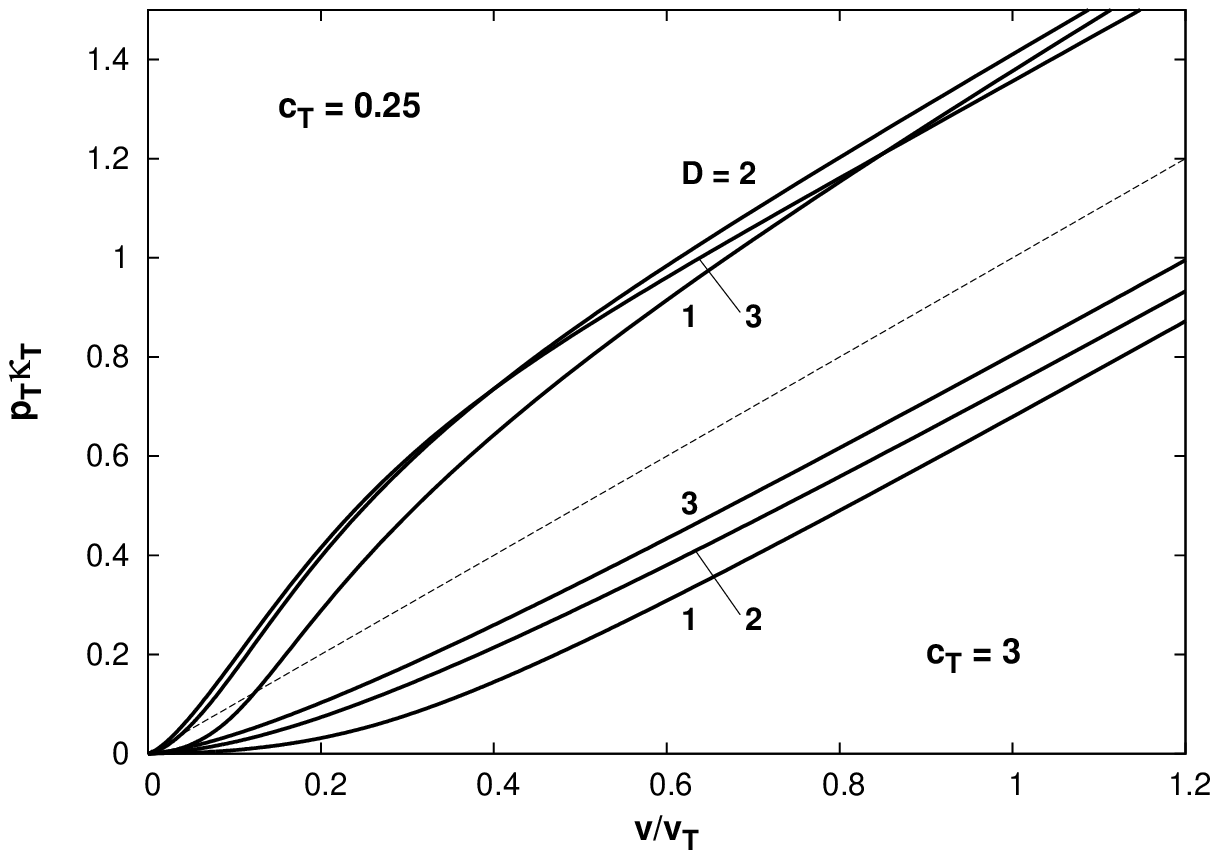}
   \caption{Isochoric heat capacity (scaled), isobaric expansivity, and
     isothermal compressibility in $D = 1,2,3$ for reduced coupling strengths
     $c_{v,p,T}=0.25$ and $c_{v,p,T}=3$. The dashed lines represents the FD
     curves for $\mathcal{D}=\infty$. To avoid cluttering we have omitted one
     expansivity curve.}
  \label{fig:seven}
\end{figure}

In Fig.~\ref{fig:seven} the focus is on the $\mathcal{D}$-dependence of the
three response functions at $c_{v,p,T}=0.25$ (weak coupling) and $c_{v,p,T}=3$
(strong coupling). The goal is to gain further insight into how gases with
boson-like, fermion-like, and crossover features in $\mathcal{D}=1,2,3$
evolve into one and the same FD system as $\mathcal{D}\to\infty$ with an emergent
singularity at $T/T_v=T/T_p=e^{-1}$ in isochoric and isobaric processes.

For the scaled heat capacity we show weak-coupling results in the main plot and
strong-coupling results in the inset. the dependence on coupling strength of
the results in $\mathcal{D}=1,2,3$ is conspicuous but becomes imperceptibly
small for $\mathcal{D}\gtrsim20$. The universal FD result for
$\mathcal{D}=\infty$ (dashed line) is a simple step function,
\begin{equation}
  \label{eq:37}
  \lim_{\mathcal{D\to\infty}}\frac{C_v}{(\mathcal{D}/2)k_B}= 
\Theta(T-T_c),\quad T_c/T_v=e^{-1}.
\end{equation}
The decrease in initial slope with increasing $\mathcal{D}$ is clearly visible
in the curves for $\mathcal{D}=1,2,3$ but overall convergence toward the step
function is slow. The scaled heat capacity of the ideal BE gas in
$\mathcal{D}=\infty$ also consists of two terms, one being a step function as
in (\ref{eq:37}) but with $T_c/T_v=1$ and the other being a $\delta$-function
representing the latent heat \cite{PMK07}.

For the isobaric expansivity the weak-coupling and strong-coupling curves are
shown in the same plot. There is very little variation between $\mathcal{D}=1$
and $\mathcal{D}=3$ for the strong-coupling case $(c_p=3)$. Somewhat larger and
more systematic variation occurs in the weak-coupling case $(c_p=0.25)$. The
universal FD result for $\mathcal{D}=\infty$ (shown dashed) is
\begin{equation}
  \label{eq:38}
  T_p\kappa_p= \frac{T_p}{T}\Theta(T-T_c),\quad T_c/T_p=e^{-1}.
\end{equation}
Convergence is slow, but evident in the curves for $\mathcal{D}=40,300$.
Comparing the result (\ref{eq:38}) for the ideal FD gas with that of the ideal
BE gas, both in $\mathcal{D}=\infty$, we find that the latter also has the form
(\ref{eq:38}) at $T>T_c$ but with $T_c/T_p=1$. The expansivity of the FD gas is
zero at $T<T_c$, in the BE gas it is undefined \cite{PMK07}.

For the isothermal compressibility the universal FD line for
$\mathcal{D}=\infty$ (dashed line),
\begin{equation}
  \label{eq:41}
p_T\kappa_T=\frac{v}{v_T},  
\end{equation}
 is indistinguishable from the MB result.
The weak-coupling and strong-coupling curves are located
on opposite sides of that line. Convergence is apparent in the
curves for $\mathcal{D}=1,2,3$ in the strong-coupling case $(c_T=3)$ but not in
the weak-coupling case $(c_T=0.25)$. The isothermal compressibility of ideal BE
gas is also described by the result (\ref{eq:41}) but only for
$v/v_T>v_c/v_T=1$. At $v/v_T<1$ the bosonic result is infinite \cite{PMK07}.

\subsection{Speed of sound}\label{sec:speso}  
\begin{figure}[htb]
  \centering
   \includegraphics[width=86mm]{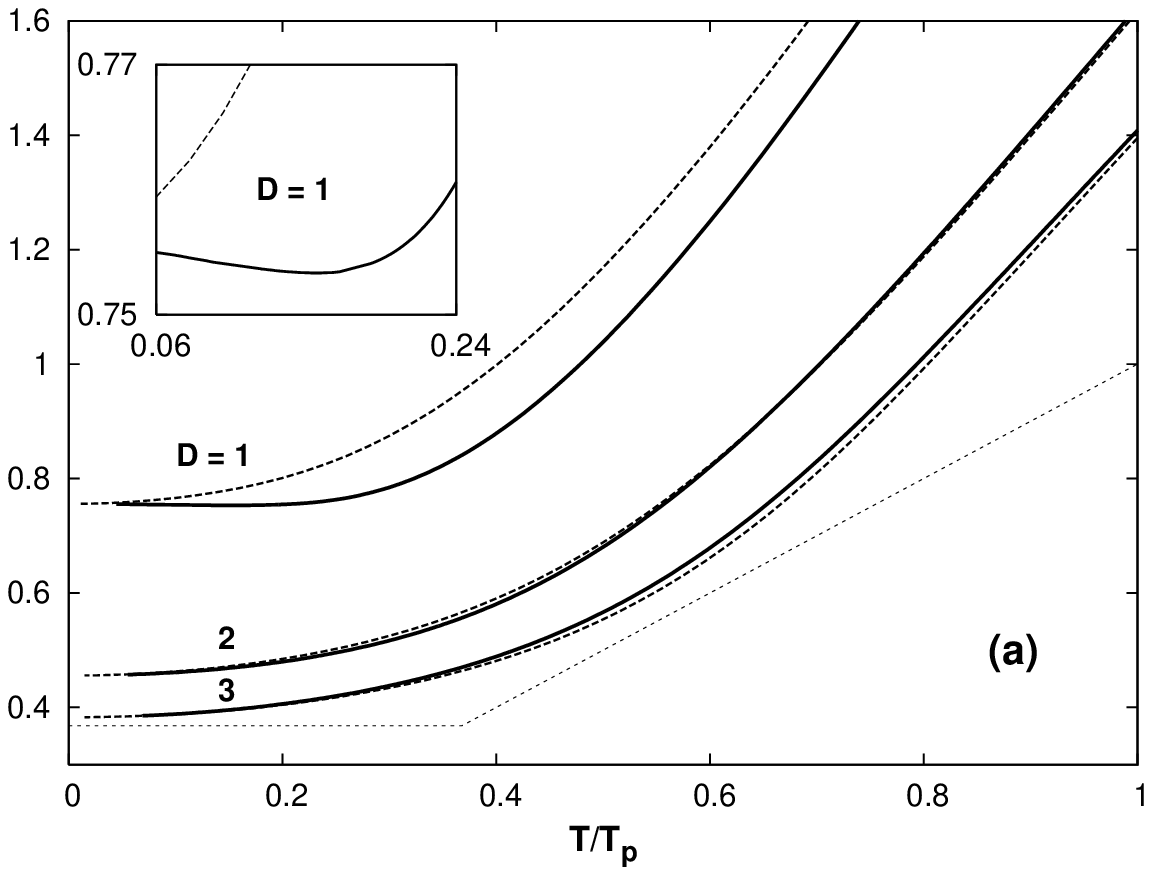}
   \includegraphics[width=86mm]{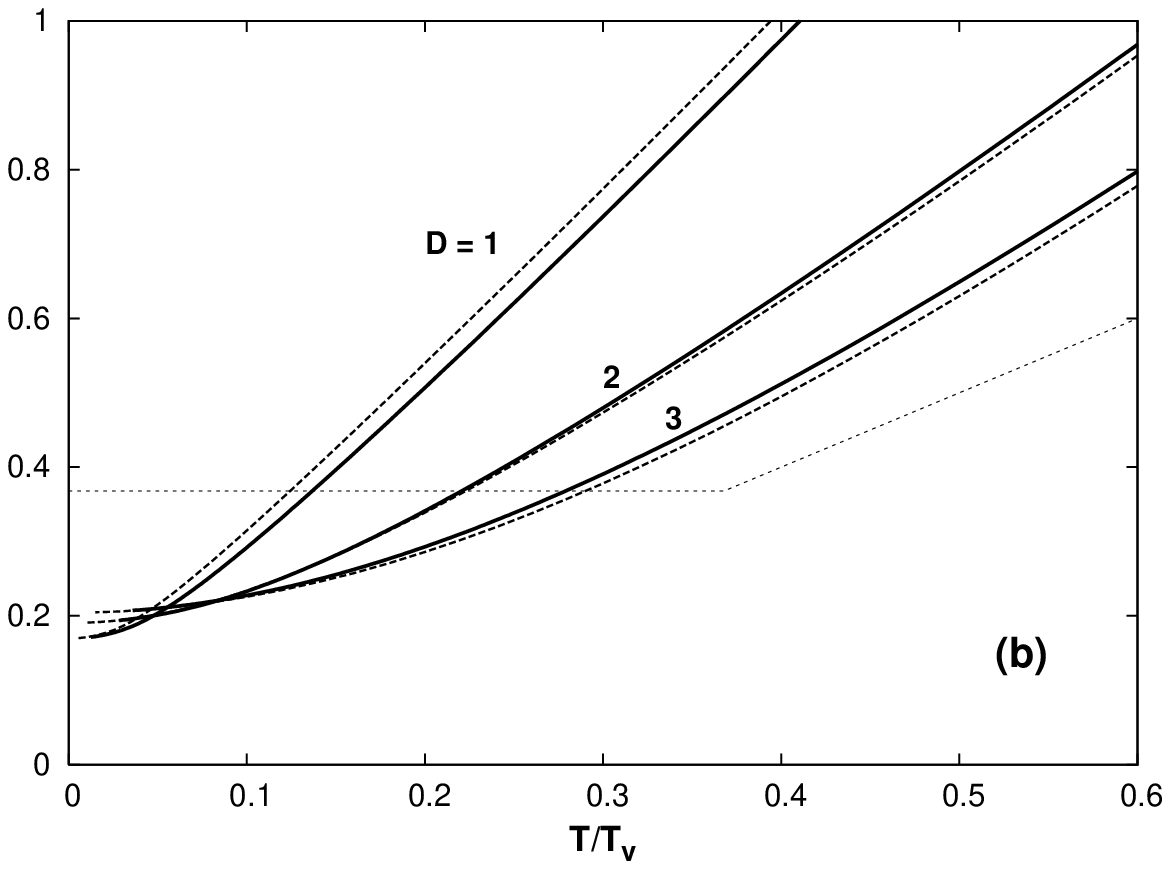}
   \caption{(a) Main plot: Speed of sound (squared and scaled), $mc^{2}/
     k_{B}T_{p}$, at constant (average) pressure versus $T/T_{p}$ in
     $\mathcal{D}=1,2,3$ (solid lines) and $\mathcal{D}=\infty$ (dotted line).
     Also shown are scaled isobars, $(1+2/ \mathcal{D})(v/v_{p})$ versus
     $T/T_{p}$ (dashed lines). Inset: Zoomed extract of the $\mathcal{D}=1$
     results near the minimum of the speed of sound data.  (b) Corresponding
     data for $mc^{2}/ k_{B}T_{v}$ and
     $(1+2/ \mathcal{D})(p/p_{v})$ versus $T/T_{v}$.  All data
     pertain to a weak-coupling situation ($c_{v,p}=0.25$).}
  \label{fig:eight}
\end{figure}

The speed of sound as inferred from $c=(\rho\kappa_S)^{-1/2}$, where $\rho=m/v$
is the mass density and $\kappa_S$ the adiabatic compressibility, can be
brought into the form \cite{fn3}
\begin{equation}
  \label{eq:39}
  \frac{mc^2}{k_BT}= \frac{(v/v_T)}{(p_T\kappa_T)}\left[1+ 
    \frac{(T/T_p)^2(v/v_T)(T_p\alpha_p)^2}{(p_T\kappa_T)(C_v/k_B)}\right],
\end{equation}
which only involves dimensionless quantities previously determined in terms of
the NLS functions,
\begin{eqnarray}
  \label{eq:40}
  \frac{mc^2}{k_BT}&=& 
\frac{\frac{\partial}{\partial z}F_{p}^{(\mathcal{D})}(z,\bar{c})}
  {\frac{\partial}{\partial z}F_{\mathcal{N}}^{(\mathcal{D})}(z,\bar{c})}
  \left\{1+ \frac{2}{\mathcal{D}}
\frac{\frac{\partial}{\partial z}F_{p}^{(\mathcal{D})}(z,\bar{c})}
  {\frac{\partial}{\partial z}F_{U}^{(\mathcal{D})}(z,\bar{c})}~\right.
\nonumber \\
  &&\hspace{-8mm}\times\left.\frac{\left[\left(\frac{\mathcal{D}}{2} +1\right)
  \frac{F_p^{(\mathcal{D})}(z,\bar{c})
\frac{\partial}{\partial z}F_\mathcal{N}^{(\mathcal{D})}(z,\bar{c})}
  {F_\mathcal{N}^{(\mathcal{D})}(z,\bar{c})
\frac{\partial}{\partial z}F_p^{(\mathcal{D})}(z,\bar{c})} - 
  \frac{\mathcal{D}}{2}\right]^2}{\left[\left(\frac{\mathcal{D}}{2} +1\right)
  \frac{F_U^{(\mathcal{D})}(z,\bar{c})
\frac{\partial}{\partial z}F_\mathcal{N}^{(\mathcal{D})}(z,\bar{c})}
  {F_\mathcal{N}^{(\mathcal{D})}(z,\bar{c})
\frac{\partial}{\partial z}F_U^{(\mathcal{D})}(z,\bar{c})} - 
  \frac{\mathcal{D}}{2}\right]}\right\}.
\end{eqnarray}
Here $\bar{c}$ must be replaced by $c_T$, $c_vx$, or $c_py$ depending on
whether we are considering an isothermal, isochoric, or isobaric process,
respectively. In Fig.~\ref{fig:eight} we present data for the $T$-dependence of
the speed of sound of a weak-coupling system under isobaric and isochoric
conditions.

It is well-known that in ideal gases the curves for $mc^2/k_BT$ differ from
those of the isobars or isochores only by a multiplicative factor $(1+2/
\mathcal{D})$. We have seen that this relation still holds in the presence of
fractional statistics \cite{PMK07}. The data shown here for the
generalized NLS model demonstrate that no such relation holds any longer in the
presence of a statistical interaction that is not reducible to a simple
exclusion principle.

The deviations appear to be strongest in $\mathcal{D}=1$. All deviations are
expected to fade away in the limt $\mathcal{D}\to\infty$ when ideal gas behavior is
restored as explained in Sec.~\ref{sec:pddi}. Particularly noteworthy is the
observation that the $T$-dependence of the speed of sound in $\mathcal{D}=1$ at
constant (average) pressure undergoes a minimum as highlighted in the inset. No
such minimum exists in the isobar. 

We attribute this effect to the crossover between boson-like features at
high $T$ and fermion-like features at low $T$. The general trend, realized in
ideal gases is that the speed of sound decreases monotonically upon
cooling. Superimposed on this is another trend that signals softness when
boson-like features are predominant and stiffness when fermion-like features
are predominant. 

For stronger coupling (e.g. $c_{v.p}=3$) the deviations of the speed-of-sound
data from the scaled isobar or isochore are of a similar kind and size. We have
detected no minimum at $T>0$ in these data. If such a minimum exists at all it
must occur at very low $T$, out of reach of our numerical analysis.

%
\section{Conclusion}\label{sec:concl}  
%

The exact thermodynamic analysis of the generalized NLS model, a quantum gas in
$\mathcal{D}$ dimensions with a statistical two-body interaction, has yielded
significant deviations from characteristic ideal quantum gas behavior in
several respects. 

For given coupling strength $0<c<\infty$ (i) the average level occupancy $\langle
n(k)\rangle$ is no longer a unique function of $(k^{2}-\mu)/k_{B}T$ and independent
of $\mathcal{D}$; (ii) the quantities $p\lambda_{T}^{\mathcal{D}}/k_{B}T$,
$\mathcal{N}\lambda_{T}^{\mathcal{D}}/V$, and
$(U\lambda_{T}^{\mathcal{D}}/V)/(k_{B}T\mathcal{D}/2)$ are no longer unique
functions of the fugacity $z$; (iii) the two quantities
$p\lambda_{T}^{\mathcal{D}}/k_{B}T$ and
$(U\lambda_{T}^{\mathcal{D}}/V)/(k_{B}T\mathcal{D}/2)$ are no longer identical.

Among the consequences are (i) that the $T$-dependence of the internal
energy is no longer of the same shape as the isochore; (ii) that the quantity
$pV/ \mathcal{N}k_{B}T$ is no longer a function of $z$ alone in given
$\mathcal{D}$; (iii) that there is no longer any simple relation between the
speed of sound and the isochore or isobar.

In any finite $\mathcal{D}$ the statistical interaction of the generalized NLS
model smoothly interpolates between an ideal BE gas in the weak-coupling limit
$(c=0)$ and an ideal FD gas in the strong-coupling limit $(c=\infty)$. In the
limit $\mathcal{D}\to\infty$ the system behaves like an ideal BE gas for $c=0$
and like an ideal FD gas for $c>0$. In $\mathcal{D}=\infty$ both quantum gases
feature a phase transition at $T_{c}>0$ along isochores or isobars. The
transition is of first order in the BE case and of second order in the FD case.

\vspace*{-5mm}
%
\acknowledgments
%
Financial support from the DFG Schwerpunkt \textit{Kollektive Quantenzust{\"{a}}nde
  in elektronischen 1D {\"{U}}bergangsmetallverbindungen} (for M.K.)  is gratefully
acknowledged. We have greatly benefited from discussions with
Prof. A. E. Meyerovich.  


\begin{thebibliography}{100}
\bibitem{fn4}
The same argument applies to the case $\bar{c}=0$ but only for $z\leq
1$. Indeed we have $f_\infty(z)=g_{\infty}(z)=z$ for $0\leq z\leq 1$.
\bibitem{fn3} 
For consistency, we must use reference values $T_{v}$, $T_{p}$,
  and $v_{T}$ derived from the first expression (\ref{eq:11}) for $\lambda_{T}$ in
  this context.
\end{thebibliography}
\end{document}